\newcommand*{\ATLASLATEXPATH}{latex/}

\pdfoutput=1 
\documentclass[cernpreprint,texlive=2011,txfonts,UKenglish]{\ATLASLATEXPATH atlasdoc}

\usepackage[block=space]{\ATLASLATEXPATH atlaspackage}
\usepackage{\ATLASLATEXPATH atlasbiblatex}
\usepackage{\ATLASLATEXPATH atlascontribute}
\usepackage{\ATLASLATEXPATH atlasphysics}

\usepackage{url}
\usepackage[utf8]{inputenc}

\graphicspath{{logos/}{figures/}}

\usepackage{paper-defs}
\usepackage{paper-results}

\AtlasTitle{Search for anomalous couplings in the \Wtb\ vertex from the measurement
of double differential angular decay rates of single top quarks
produced in the \tch\ with the ATLAS detector}

\author{The ATLAS Collaboration}

\date{\today}

\PreprintIdNumber{CERN-PH-EP-2015-236}

\AtlasJournal{JHEP}

\AtlasAbstract{%
  The electroweak production and subsequent decay of single top quarks 
  is determined by the properties of the \Wtb\ vertex. This vertex can 
  be described by the complex parameters of an 
  effective Lagrangian. An analysis of angular distributions of the decay 
  products of single top quarks produced in the \tch\ constrains these 
  parameters simultaneously. The analysis described in this paper uses 
  \lumi\ of proton--proton collision data at \cmenergy\ collected with 
  the ATLAS detector at the LHC. Two parameters are measured simultaneously 
  in this analysis. The fraction $\fu$ of decays containing transversely 
  polarised $W$ bosons is measured to be $\cvfucombunc$. The phase $\delm$ 
  between amplitudes for transversely and longitudinally polarised $W$ bosons 
  recoiling against left-handed $b$-quarks is measured to be $\cvdelmcombunc$. 
  The correlation in the measurement of these parameters is $\corrparamfull$. 
  These values result in two-dimensional limits at the 95\% confidence level 
  on the ratio of the complex coupling parameters \gr\ and \vl, yielding 
  $\regrvl\in\regrvllimits$ and $\imgrvl\in\imgrvllimits$ with a correlation 
  of $\corrcoupfull$. The results are in good agreement with the predictions 
  of the Standard Model.
}

\hypersetup{pdftitle={ATLAS draft},pdfauthor={The ATLAS Collaboration}}

\begin{document}

\maketitle

\section{Introduction}
\label{sec:intro}

The top quark is the heaviest known fundamental particle, making the measurement of its production and decay kinematics a unique probe of physical processes beyond 
the Standard Model (SM). The top quark was first observed in 1995 at the Tevatron~\cite{CDFTopDiscovery, D0TopDiscovery} through its dominant production channel, 
top-quark pair (\ttbar) production via the flavour-conserving strong interaction. An alternative process produces single top quarks through the weak interaction, 
first observed at the Tevatron in 2009~\cite{CDFSgTop,D0SgTop}.

\begin{figure}[!htb]
  \centering
  \centerline{\makebox{
      \subfloat[]{\includegraphics[width=2.4in]{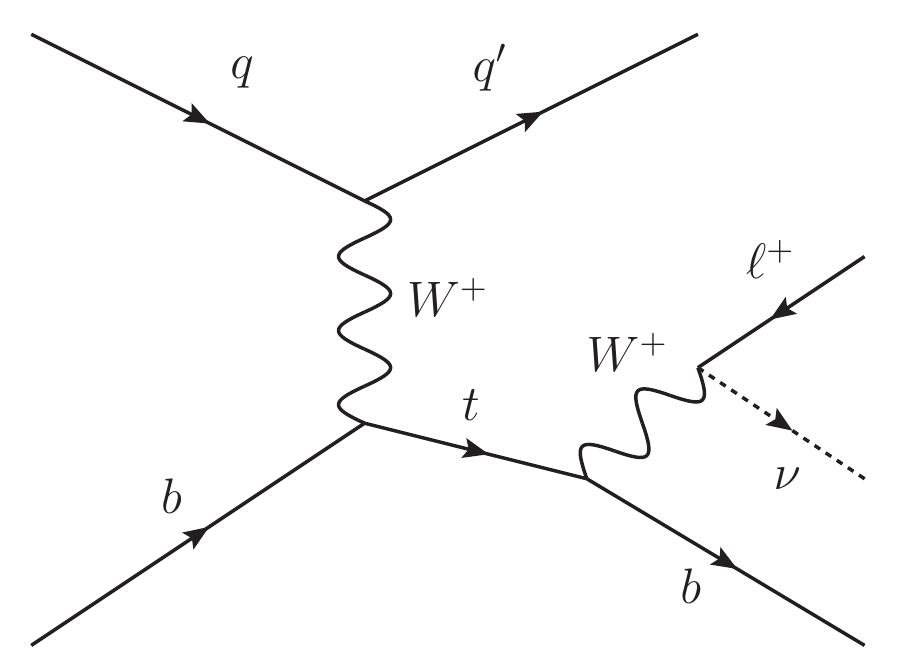}}
      \subfloat[]{\includegraphics[width=2.4in]{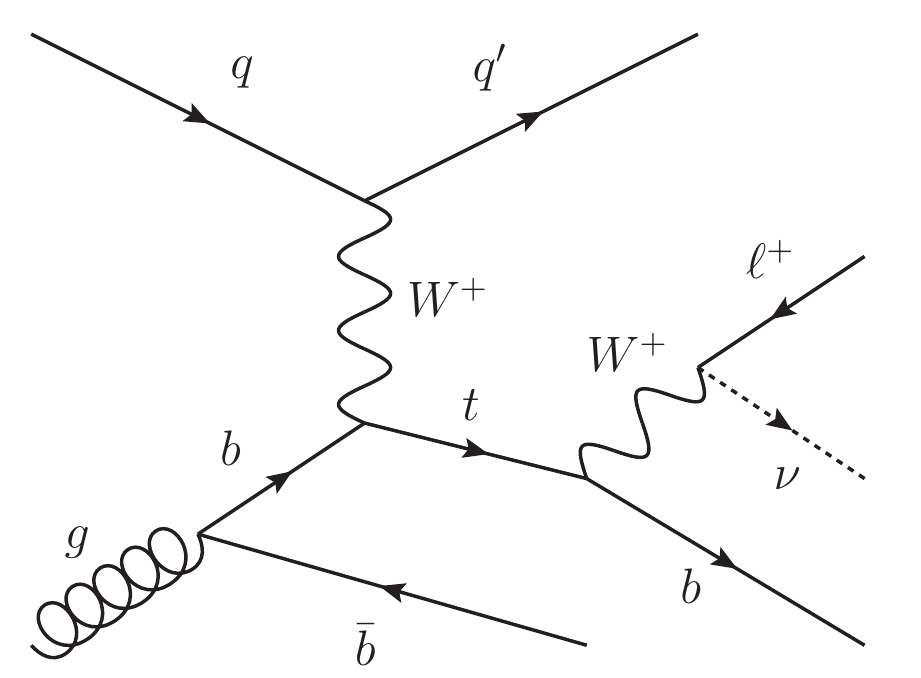}}
  }}
  \caption{ Representative Feynman diagrams for \tch\ single top-quark production and decay. Here $q$ represents a $u$ or $\bar{d}$ quark, and $q^\prime$ represents 
            a $d$ or $\bar{u}$ quark, respectively. The initial $b$-quark arises from (a) a sea $b$-quark in the $2\to2$ process, or (b) a gluon splitting into a 
            $b\bar{b}$ pair in the $2\to3$ process.}
  \label{fig:tChannelDiagram}
\end{figure}

The \tch\ exchange of a virtual $W$ boson is the dominant process contributing to single top-quark production (see \Figure{fig:tChannelDiagram}). The production 
cross-section is calculated for proton--proton (\textit{pp}) collisions at \cmenergy\ using a next-to-leading-order (NLO) QCD prediction with resummed 
next-to-next-to-leading logarithmic (NNLL) accuracy, referred to as approximate next-to-next-to-leading order (NNLO). With a top-quark mass of 172.5~\GeV\ and 
\textsc{MSTW2008NNLO}~\cite{Martin:2009iq} parton distribution function (PDF) sets, the cross-section for the \tch\ process is calculated to be 
64.6$^{+2.6}_{-1.7}$~pb~\cite{Kidonakis:2011wy}. The uncertainties correspond to the sum in quadrature of the error obtained from the MSTW PDF set at the 90\% 
confidence level (C.L.) and the factorisation and renormalisation scale uncertainties. 

New physics resulting in corrections to the \Wtb\ vertex would affect \tch\ single top-quark production and decay, and can be probed through processes which affect 
variables sensitive to the angular distributions of the final-state particles from the top-quark decay. Within the effective field theory framework, the Lagrangian 
can be expressed in full generality as~\cite{AguilarSaavedra:2008zc,AguilarSaavedra:2009mx}:
\begin{equation}
  \Lagr_{\text{\textit{Wtb}}} = - \frac{g}{\sqrt{2}}{\overline{b}}\gamma^\mu \left (\vl \ProjL + \vr \ProjR \right )tW^-_\mu - 
  \frac{g}{\sqrt{2}}{\overline{b}}\frac{i\sigma^{\mu\nu}q_{\nu}}{\mW} \left (\gl \ProjL + \gr \ProjR \right)tW^-_\mu + \text{h.c.,}
\label{eq:WtbLagr}
\end{equation}
where $g$ is the weak coupling constant, and \mW\ and $q_{\nu}$ are the mass and the four-momentum of the $W$ boson, respectively. $\ProjLR\equiv(1\mp\gamma^5)/2$ 
are the left- and right-handed projection operators and $\sigma^{\mu\nu}=i[\gamma^{\mu},\gamma^{\nu}]/2$. $\vlr$ and $\glr$ are the complex left- and right-handed 
vector and tensor couplings, respectively. The couplings could also be expressed using the Wilson coefficients~\cite{Wilson:1969zs} of the relevant dimension-six 
operators\footnote{In general the couplings can depend on the momentum transfer $q^2$. 
                   Since this analysis is mainly sensitive to
                   top-quark decays to an on-shell $W$ boson, where $q^2=m_W^2$,
                   no $q^2$ dependence is considered.}, 
described in Refs.~\cite{Buchmuller:1985jz,AguilarSaavedra:2008zc}.  In the SM at tree level, $\vl=V_{\text{\textit{tb}}}$, which is an 
element in the Cabibbo--Kobayashi--Maskawa (CKM) matrix, while the anomalous couplings $\vr=\glr=0$. Deviations from these values would provide hints of physics 
beyond the SM, and complex values could imply that top-quark decay has a $\mathcal{CP}$-violating component~\cite{AguilarSaavedra:2010nx}.

Indirect constraints on $\vr$, $\gl$, and $\gr$ were obtained~\cite{Grzadkowski:2008mf,ref:GeneralAnalysisWtb} from precision measurements of $B$-meson decays. 
These result in very tight constraints on $\vr$ and $\gl$, but much looser constraints on $\gr$. Calculations of the anomalous couplings in specific models predict a much larger contribution to $\gr$ than to $\vr$ or $\gl$ \cite{ref:BernreutherTDecayBSM}. 
However, within the models studied so far for tensor couplings, the magnitude of $\gr$ is expected to be $<0.01$~\cite{ref:BernreutherTDecayBSM,Duarte:2013zfa}.
Measurements of the $W$ boson polarisation fractions~\cite{ref:TevatronWpol,Aad:2012ky,ref:CMSWhellplusJets7TeV,ref:CMSWhelSgTop8TeV} are sensitive to the magnitude of combinations of anomalous couplings. 
In order to obtain constraints on $\regr$ in those measurements, the anomalous couplings are taken to be purely real, $\vr = \gl = 0$ is assumed and $\vl$ is fixed to 1. 
In this analysis, the anomalous couplings $\vl$ and $\gr$ are allowed to be complex, and the ratios $\regrvl$ and $\imgrvl$ are measured.

This paper is organised as follows. \Section{sec:definition} describes the coordinate system and parameterisation used in the measurement. \Section{sec:detector} 
gives a short description of the ATLAS detector, then \Section{sec:data_and_MC_samples} describes the simulated samples used to predict properties of the \tch\ 
signal and the background processes, and the data samples used. \Section{sec:selection} describes the definitions of the objects used to identify \tch\ events, and 
the set of criteria applied to reduce the number of background events. The procedures for modelling background processes are described in \Section{sec:bkg_est}, 
along with comparisons between the predictions and data. \Section{sec:analysis_method} describes the efficiency, resolution, and background models used to translate 
the distribution of true \tch\ signal events to the distribution of reconstructed signal and background events, and how the parameters of the model are estimated. 
\Section{sec:systematics} quantifies the sources of uncertainty important in this measurement, \Section{sec:result} presents the resulting central value and 
covariance matrix for the model parameters and the ratios $\regrvl$ and $\imgrvl$, and the conclusions are given in \Section{sec:conclusion}.

\section{Measurement definition}
\label{sec:definition}

\begin{figure}[!htb]
  \centering
  \includegraphics[width=2.7in]{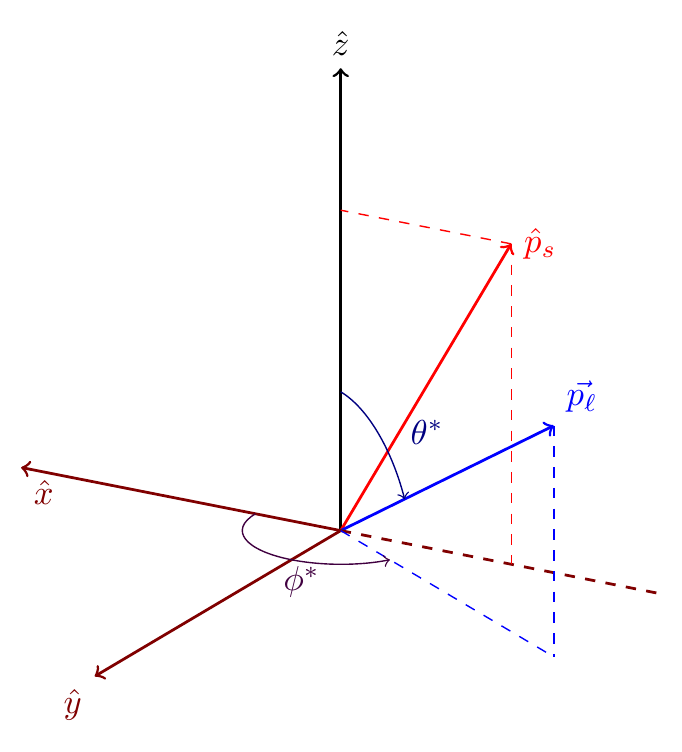}
  \caption{Definition of the coordinate system with $\hat{x}$, $\hat{y}$, and $\hat{z}$  defined as shown from the momentum directions of 
            the $W$ boson, $\Wdir \equiv \hat{z}$, and the spectator jet, $\Sdir$, in the top-quark rest frame. The angles $\thetaS$ and $\phiS$ indicate 
            the lepton direction \Ldir\ in this coordinate system.}
\label{fig:newaxes}
\end{figure}

An event-specific coordinate system is defined for analysing the decay of the top quark in its rest frame using the directions of the jets, lepton, and $W$ boson in 
the final state, depicted in \Figure{fig:newaxes}. The $\hat{z}$-axis is chosen along the direction of the $W$ boson boosted into the top-quark rest frame, 
$\hat{z}\equiv\Wdir={\Wmom}/{|\Wmom|}$. The angle $\thetaS$ between $\Wdir$ and the momentum of the lepton in the $W$ boson rest frame, $\Lmom$, is the same angle 
used to measure the $W$ boson helicity fractions in top-quark decays~\cite{Aad:2012ky,ref:CMSWhellplusJets7TeV,ref:CMSWhelSgTop8TeV,ref:CMSWhelSgTop8TeV}. The spin 
of single top quarks produced in the \tch, \topspin, is predicted to be polarised along the direction of the light quark ($q^\prime$ in \Figure{fig:tChannelDiagram}) 
in the top-quark rest frame~\cite{ref:Mahlon1996pn}, called the spectator-quark direction, $\Sdir$. If this jet defines the spin analysing direction, the degree of 
polarisation is shown in Refs.~\cite{Mahlon:1999gz,Schwienhorst:2010je} to be $P\equiv \Sdir \cdot \topspin / |\topspin| \approx 0.9$ at \cmenergy\ for SM couplings. 
A three-dimensional coordinate system is defined from the $\Wdir$--$\Sdir$ plane and the perpendicular direction, with $\hat{y}=\Sdir\times\Wdir$ and 
$\hat{x}=\hat{y}\times\Wdir$. In this coordinate system, the direction of the lepton ($\ell$) in the $W$ boson rest frame \Ldir\ is specified by $\thetaS$ and the 
complementary azimuthal angle $\phiS$. 

These angles arise as a natural choice for measuring a normalised double differential distribution, in the helicity basis (top-quark rest frame), of the leptonic 
decay of the $W$ boson arising from the top quark. In this basis, the $t\to W b$ transition is determined by four helicity amplitudes, where the $W$ boson and 
$b$-quark are either both right-handed, both left-handed, or the $W$ boson is longitudinal and the $b$-quark is either left- or right-handed. The resulting angular 
distribution of final-state leptons in $\thetaS$ and $\phiS$ is expressed~\cite{Boudreau:2013yna} as a series in spherical harmonics, $Y_l^m (\thetaS,\phiS)$, parameterised by 
$\params = \left(\fu,\fup,\fzp,\delp,\delm\right)$ and $P$:
\begin{equation}
  \begin{split}
    \rho(\thetaS,\phiS;&\ \params,P) \equiv \frac{1}{N} \frac{\mathrm{d}N}{\mathrm{d}\Omega^*} = \sum\limits_{l=0}^{2} \sum\limits_{m=-l}^{l} a_{l,m}(\params,P) Y_l^m (\thetaS,\phiS), 
    \text{\ with\ } \\ 
    a_{0,0} &=\frac{1}{\sqrt{4\pi}}, \qquad 
    a_{1,0}=\frac{\sqrt{3}}{\sqrt{4\pi}}f_1\left(f_{1}^{+}-\frac{1}{2}\right), \qquad
    a_{2,0}=\frac{1}{\sqrt{20\pi}}\left(\frac{3}{2}f_1-1\right), \\
    a_{1,1} &= -a_{1,-1}^* = P \frac{\sqrt{3\pi}}{16} \sqrt{f_1(1-f_1)} \left\{\sqrt{f_1^+ f_0^+} e^{i\delta_+} + \sqrt{(1-f_1^+)(1-f_0^+)} e^{-i\delta_-}\right\}, \\
    a_{2,1} &= -a_{2,-1}^* = P \frac{\sqrt{3\pi}}{16\sqrt{5}} \sqrt{f_1(1-f_1)} \left\{ \sqrt{f_1^+ f_0^+} e^{i\delta_+} - \sqrt{(1-f_1^+)(1-f_0^+)} e^{-i\delta_-} \right\}, \\
    a_{2,2} &= a_{2,-2} = 0.
  \end{split}
  \label{eq:signal_model}
\end{equation}
The restriction to $l \le 2$ in \Equation{eq:signal_model} is caused by the limited spin states of the initial and final state fermions and the vector boson at the weak vertex.

The parameters $\params$ describing this distribution can be expressed at tree level in terms of four linear combinations of the anomalous couplings. The fractions, 
$\fu$, $\fup$, and $\fzp$, depend on the magnitude of these four combinations, while $\delp$ and $\delm$ depend on the phases between pairs of combinations. In 
addition there is a small correction due to the finite mass of the $b$-quark, $m_b$. Defining $x_W=\mW/m_t$ and $x_b=m_b/m_t$, where $m_t$ is the top-quark mass, 
the parameters are given by:
\begin{itemize}
\item $\fu$, the fraction of decays containing transversely polarised $W$ bosons,
  \begin{equation}
    \fu =\frac{2\left(|\xw\vl-\gr|^2+|\xw\vr-\gl|^2\right)+\mathcal{O}(\xb)}
              {2\left(|\xw\vl-\gr|^2+|\xw\vr-\gl|^2\right)+|\vl-\xw\gr|^2+|\vr-\xw\gl|^2+\mathcal{O}(\xb)} \\
  \end{equation}
\item $\fup$, the fraction of transversely polarised $W$ boson decays that are right-handed,
  \begin{equation}
    \fup = \frac{|\xw\vr-\gl|^2+\mathcal{O}(\xb)}{|\xw\vl-\gr|^2+|\xw\vr-\gl|^2+\mathcal{O}(\xb)} \\
  \end{equation}
\item $\fzp$, in events with longitudinally polarised $W$ bosons, the fraction of $b$-quarks that are right-handed,
  \begin{equation}
    \fzp  = \frac{|\vr-\xw\gl|^2+\mathcal{O}(\xb)}{|\vr-\xw\gl|^2+|\vl-\xw\gr|^2+\mathcal{O}(\xb)}\\
  \end{equation}
\item $\delp$, the phase between amplitudes for longitudinally polarised and transversely polarised $W$ bosons recoiling against right-handed $b$-quarks,
  \begin{equation}
    \delp = \arg\left((\xw\vr-\gl){(\vr-\xw\gl)}^*+\mathcal{O}(\xb)\right)\\
  \end{equation}
\item $\delm$, the phase between amplitudes for longitudinally polarised and transversely polarised $W$ bosons recoiling against left-handed $b$-quarks,
  \begin{equation}
    \delm = \arg\left((\xw\vl-\gr)(\vl-\xw\gr)^*+\mathcal{O}(\xb)\right)
  \end{equation}
\item $P$, which is considered separately from $\params$ because it depends on the production of the top quark, rather than the decay. There is no analytical 
      expression for $P$ in terms of anomalous couplings, but a parameterisation is determined in Ref.~\cite{ref:ASPolPaper} by fitting simulated samples produced 
      with the leading-order (LO) \protos~\cite{AguilarSaavedra:2008gt} 
      generator\footnote{\protos\ (PROgram for TOp Simulations) is a generator for studying 
                         new physics processes involving the top quark. It has generators for 
                         single top-quark and top-quark pair production with anomalous \Wtb\ 
                         couplings.} 
      with different values for the various couplings.
\end{itemize}
Through these expressions, measurements of the parameterisation variables, ($\params,P)$, can be used to set limits on the values of the couplings \vlr\ and \glr.

\begin{figure}
  \centering
  \centerline{\makebox{
      \subfloat[]{\includegraphics[width=\textwidth]{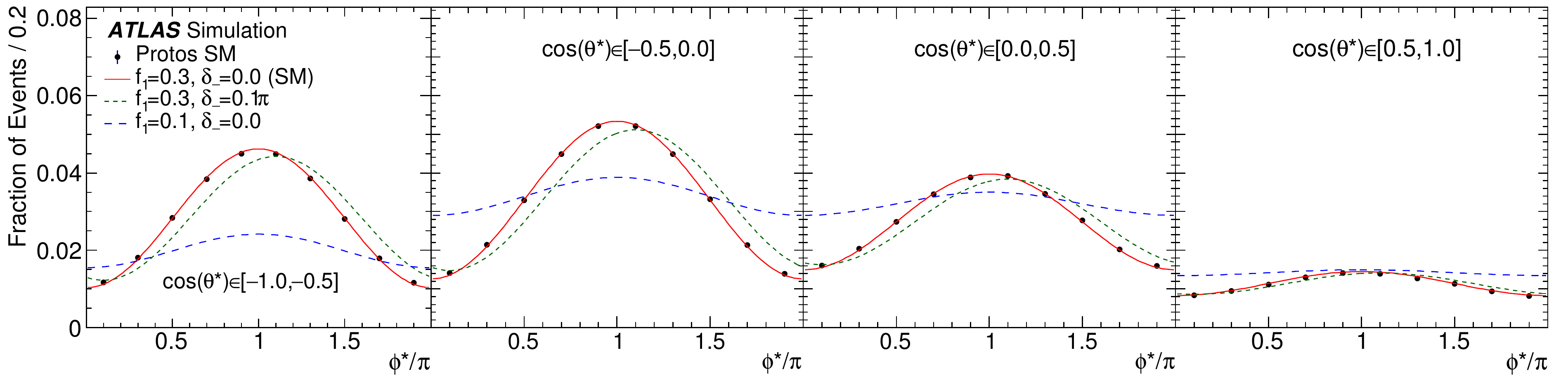}}
  }}
   \centerline{\makebox{
      \subfloat[]{\includegraphics[width=3.2in]{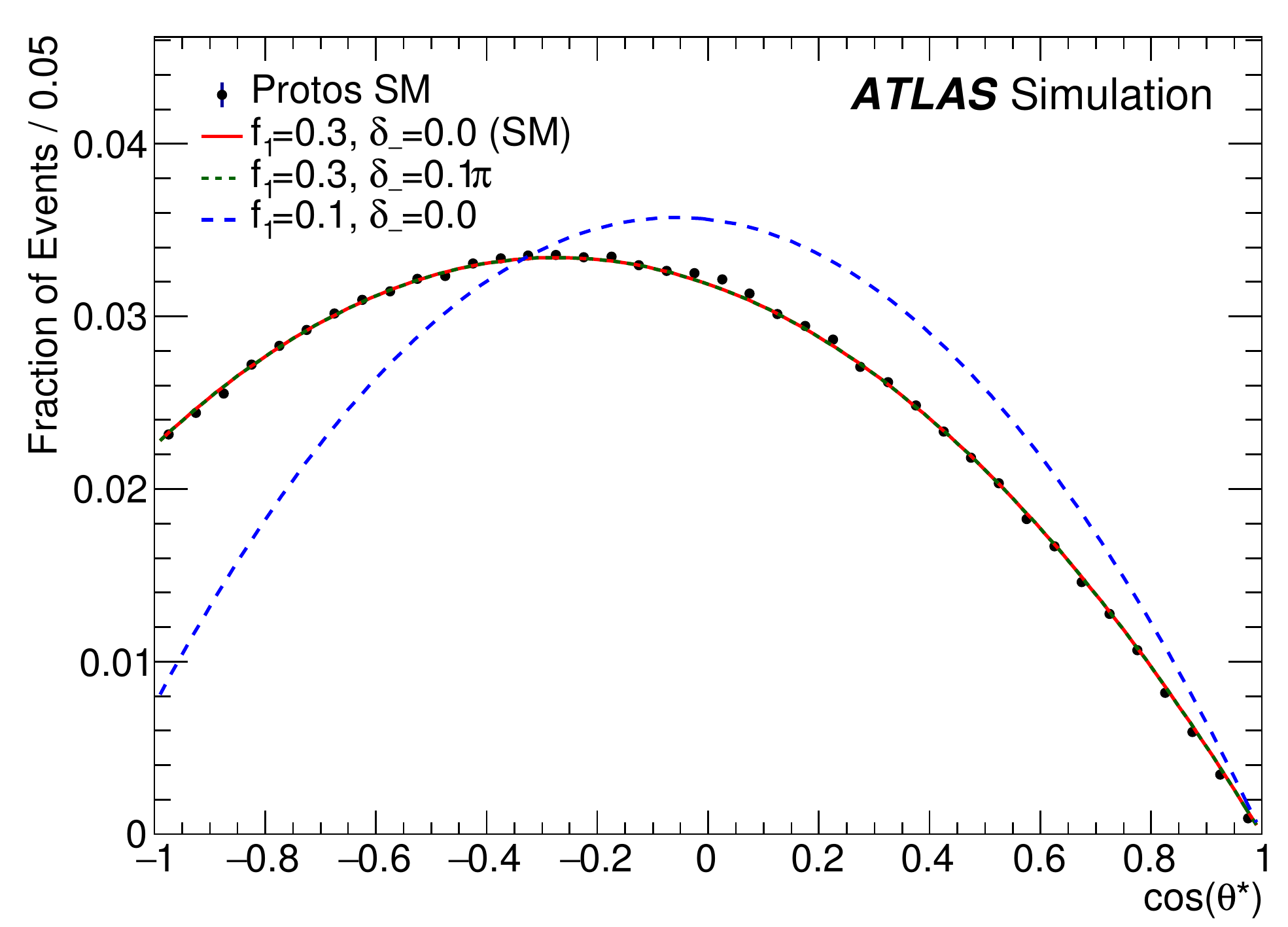}}
      \subfloat[]{\includegraphics[width=3.2in]{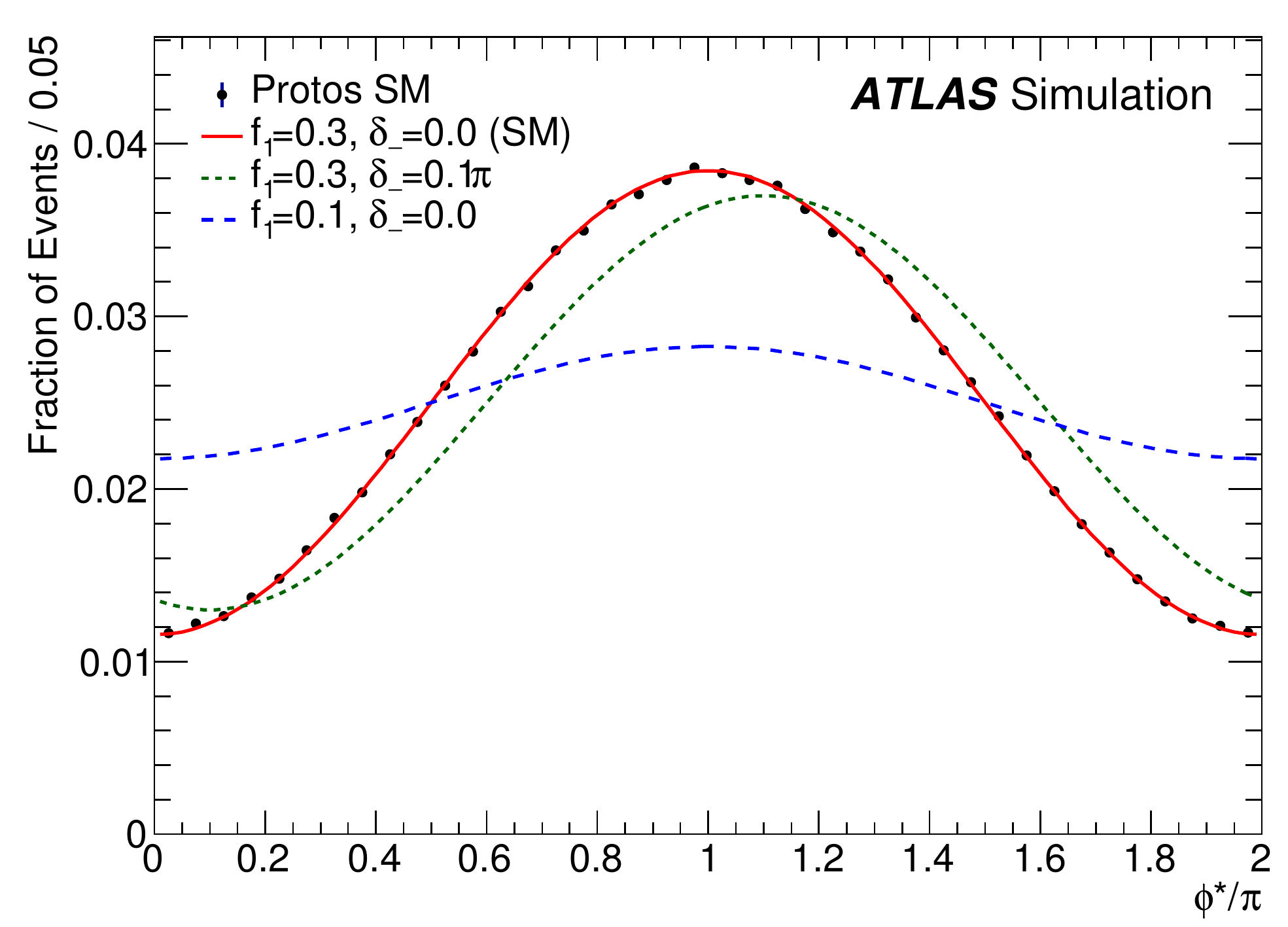}}
  }}
  \caption{ Projections into (a) $\phiS$ in bins of $\cosThetaS$, (b) $\cosThetaS$, and (c) $\phiS$ in \Equation{eq:signal_model}, for different values of $\fu$ and 
            $\delm$. The black points represent the \protos\ \tch\ signal generated with SM parameters, and the curves shown represent the signal model. For the 
            three curves shown, the parameters $\fu$ and $\delm$ are set to their values in the SM, $\fu=0.3$, $\delm=0$ (solid red), and to two sets of 
            beyond-the-SM values, $\fu=0.1$, $\delm=0$ (dashed blue) and $\fu=0.3$, $\delm=0.1\pi$ (dotted green). }
\label{fig:genModelVariation}
\end{figure}

The parameters to which this analysis is most sensitive are the fraction $\fu\in[0,1]$ and the phase $\delm\in[-\pi,\pi]$. The parameter $\fu$ can be related to the 
$W$ boson helicity fractions via $\fu=\Fp+\Fm$, where $\Fp=\fu\fup$ and $\Fm=\fu(1-\fup)$. 
Since $\fup$ and $\fzp$ are small near the SM point, the term in \Equation{eq:signal_model} that is proportional to $e^{i\delta_+}$ is nearly zero.
Thus $\delta_+$ cannot be constrained and $\fzp$ cannot be separated from $P$.
Constraints on $\fup$ are currently better determined from limits on $\Fp$~\cite{Aad:2012ky,ref:CMSWhellplusJets7TeV}.
The variations that the parameters $\fu$ and $\delm$ induce in the model 
are demonstrated in \Figure{fig:genModelVariation} for \tch\ signal events generated with \protos. For regions of the parameter space $(\fu,\delm)$ close to the SM, 
only these parameters can be significantly constrained. The dependence of $\delm$ on $\vr$ and $\gl$ is suppressed by $x_b$, while both $\fu$ and $\delm$ are 
dependent on $\vl$ and $\gr$ directly or through $x_W$. Thus, to simplify the analysis, only variations in $\vl$ and $\gr$ are considered, while $\vr$ and $\gl$ are 
assumed to be zero. With this assumption, the values of $\fup$ and $\fzp$ are very small. The value of $P$ is also determined from the values of $\vl$ and $\gr$. 
The highest-order dependence of $\fu$ and $\delm$ on the couplings $\vl$ and $\gr$ appear as the ratio $\frac{\gr}{\vl}$, where the real and imaginary parts of this 
ratio are measured separately. This motivates quoting the results in both the parameter space $(\fu,\delm)$ and the coupling space $(\regrvl,\imgrvl)$.

\section{The ATLAS detector}
\label{sec:detector}
The ATLAS detector~\cite{Aad:2008zzm} consists of a set of sub-detector systems, cylindrical in the central region and planar in the two 
end-cap regions, that cover almost the full solid angle around the interaction 
point\footnote{ATLAS uses a right-handed coordinate system with 
  its origin at the nominal interaction point (IP) in the centre of the detector 
  and the $z$-axis along the beam pipe. The $x$-axis points from the IP to 
  the centre of the LHC ring, and the $y$-axis points upward. Cylindrical 
  coordinates $(r,\phi)$ are used in the transverse plane, $\phi$ being the 
  azimuthal angle around the beam pipe. The pseudorapidity is defined in 
  terms of the polar angle $\theta$ as $\eta=-\ln\tan(\theta/2)$.}.
ATLAS is composed of an inner tracking detector (ID) close to the interaction point, surrounded by a superconducting solenoid providing a 2~T axial magnetic field, 
electromagnetic and hadronic calorimeters, and a muon spectrometer (MS). The ID consists of a silicon pixel detector, a silicon microstrip detector (SCT), providing 
tracking information within pseudorapidity $|\eta| < 2.5$, and a straw-tube transition radiation tracker (TRT) that covers $|\eta| < 2.0$. The central 
electromagnetic calorimeter is a lead and liquid-argon (LAr) sampling calorimeter with high granularity, and is divided into a barrel region that covers 
$|\eta| < 1.475$ and end-cap regions that cover $1.375 < |\eta| < 3.2$. An iron/scintillator tile calorimeter provides hadronic energy measurements in the central 
pseudorapidity range. The forward regions are instrumented with LAr calorimeters covering $3.1 < |\eta| < 4.9$ for both the electromagnetic and hadronic energy 
measurements. The MS covers $|\eta| < 2.7$ and consists of three large superconducting toroid magnets with eight coils each, a system of trigger chambers, and 
precision tracking chambers. The ATLAS detector has a three-level trigger system~\cite{Aad:2012xs}, used to select events to be recorded for offline analysis. The 
first-level trigger is hardware-based and uses a subset of the detector information to reduce the physical event rate from 40~MHz to at most 75~kHz. The second and 
third levels are software-based and together reduce the event rate to about 300~Hz.

\section{Data and simulated samples}
\label{sec:data_and_MC_samples}

This analysis is performed using \textit{pp} collision data delivered by the LHC~\cite{LHCMachine} in 2011 at \cmenergy\ and recorded by the ATLAS experiment. 
Stringent detector and data quality requirements are applied, resulting in a data sample corresponding to a total integrated luminosity of 
\lumiAndUncInFB~\cite{ref:AtlasLumiPaper2013}. The events are selected by un-prescaled single-lepton triggers~\cite{Aad:2012xs, Aad:2014sca} that require a minimum 
transverse energy, $\ET$, of 20~\GeV\ or 22~\GeV\ for electrons and a minimum transverse momentum, \pT, of 18~\GeV\ for muons, depending on the data-taking 
conditions.

Samples of events generated using Monte Carlo (MC) simulations are produced for \tch\ signal and the background processes, and are used to evaluate models of 
efficiency and resolution, and to estimate systematic uncertainties.

Samples of simulated \tch\ single top-quark events are produced with the \acermc\ multi-leg LO generator~\cite{SAMPLES-ACER} (version 3.8) using the LO 
CTEQ6L1~\cite{Pumplin:2002vw} PDF sets. \acermc\ incorporates both the $2\to2$ and $2\rightarrow3$ processes (see \Figure{fig:tChannelDiagram}) and features an 
automated procedure to remove the overlap in phase space between them~\cite{Kersevan:2006fq}. The factorisation and renormalisation scales are set to 
\mbox{$\mu_{\text{F}} = \mu_{\text{R}}=172.5$~\GeV}. Additional \tch\ samples with different anomalous couplings are produced with 
\protos~\cite{AguilarSaavedra:2008gt} (version 2.2) using the CTEQ6L1 PDF sets. Events are generated using \protos\ with the four-flavour 
scheme\footnote{In the four-flavour scheme, the PDFs only 
                contain the quarks lighter than the $b$-quark.}, 
incorporating only the $2\to3$ process, and anomalous couplings are enabled in both the production and the decay vertices, varying $\revl$ and $\imgr$ 
simultaneously to keep the top-quark width invariant. The factorisation scale is set to $\mu_{\text{F}}^2 = - p^2_W$ for the light quark and 
$\mu_{\text{F}}^2 = p^2_{\bar{b}} + m^2_b$ for the gluon. These samples are used to evaluate \tch\ generator modelling uncertainties as described in 
\Section{sec:syst_generators}. They are also used to derive folding models with non-SM values of the couplings, used for performing validation tests in MC 
simulation and measurements in real data, as described in \Section{sec:analysis_method}. The parton showering (PS), hadronisation and underlying-event (UE) 
modelling in these samples are simulated with \pythia~\cite{Sjostrand:2006za} (version 6.426) using the Perugia 2011C set of tuned parameters (P2011C 
tune)~\cite{Skands:2010ak} with the CTEQ6L1~\cite{Pumplin:2002vw} PDF sets.

Samples of \ttbar events, \sch\ single top-quark events, and associated production of a $W$ boson and top quark ($Wt$) are produced using the \powhegbox\ NLO 
generator (version 1.0) coupled with the NLO CT10~\cite{Lai:2010vv} set of PDFs and interfaced with \pythia\ for showering and hadronisation using the P2011C tune 
with CTEQ6L1 PDF sets. Additional \ttbar\ samples are produced with \protos\ (version 2.2) using the CTEQ6L1 PDF sets. These samples are used to evaluate the 
background model with non-SM values of the couplings. The PS, hadronisation and UE in these samples are simulated with \pythia\ (version 6.426) using the AUET2B 
tune~\cite{ATLAS:2011zja} with the MRST LO**~\cite{Sherstnev:2007nd} PDF sets. All processes involving top quarks are produced assuming $m_t=172.5$~\GeV.

Vector-boson production in association with jets ($V$+jets) is simulated using the multi-leg LO \alpgen\ generator~\cite{Mangano:2002ea} (version 2.13) using 
CTEQ6L1 PDF sets and interfaced to \herwig~\cite{Corcella:2000bw} (version 6.5.20) together with the \jimmy\ UE model~\cite{SAMPLES-JIMMY} (version 4.31). The 
contributions of $W$+light-jets and $W$+heavy-jets ($W$+$b\overline{b}, W$+$c\overline{c}, W$+$c$) are simulated separately. To remove overlaps between the $n$ and 
$n + 1$ parton samples the MLM matching scheme~\cite{Mangano:2002ea} is used. Double counting between the inclusive $W + n$ parton samples and samples with 
associated heavy-quark pair production is removed using an overlap-removal algorithm based on parton--jet $\Delta R$ matching~\cite{Mangano:2006rw}. Diboson 
processes ($WW$, $WZ$ and $ZZ$) are produced using the \herwig\ generator with the MRST LO** PDF sets.

After the event-generation step, the samples are passed through the full simulation of the ATLAS detector~\cite{Aad:2010ah} based on 
\GEANT{}4~\cite{Agostinelli:2002hh}. Additional \protos\ samples are passed through the \texttt{ATLFAST2} simulation~\cite{Aad:2010ah,ATLAS:2010bfa} of the ATLAS 
detector, which uses a fast simulation for the calorimeters and their response. They are then reconstructed using the same procedure as for data. The simulation 
includes the effect of multiple \textit{pp} collisions per bunch crossing (pile-up). The events are weighted such that the distribution of the average number of 
collisions per bunch crossing is the same as in data.

\section{Object definitions and event selection}
\label{sec:selection}

The definitions of objects used in the analysis, including electrons, muons, jets and $b$-tagged jets, and missing transverse momentum, as well as the basic event 
selection are chosen to be identical to those used for the \tch\ cross-section measurements in Ref.~\cite{Aad:2014fwa}. This analysis requires exactly one isolated 
charged light lepton (electron or muon) with transverse momentum $\pT > 25$~\GeV\ and pseudorapidity $|\eta|<2.5$. Exactly one $b$-tagged jet with $|\eta| < 2.5$ 
and exactly one untagged jet with $|\eta| < 4.5$ are required, both with $\pT>30$~\GeV. The second $b$-quark coming from gluon splitting as shown in 
\Figure{fig:tChannelDiagram}(b) can result in an additional $b$-tagged jet. This second $b$-tagged jet generally has a softer $\pT$ spectrum and a broader $\eta$ 
distribution compared to the $b$-tagged jet produced in the top-quark decay. It is often not detected in the experiment and is thus not required in the event 
selection. To reject jets from pile-up collisions, the jet-vertex fraction $\epsilon_\mathrm{JVF}$, defined as the ratio of the $\sum\pT$ of all tracks in the jet 
originating at the primary vertex to the $\sum\pT$ of all tracks associated with the jet, is required to satisfy $\epsilon_\mathrm{JVF}>0.75$. The magnitude of the 
missing transverse momentum must be $\MET > 30$~\GeV. Two additional \qcd\ background rejection criteria are applied. The transverse mass of the lepton--\MET system,
\begin{displaymath}
  \mtletmiss=\sqrt{2\pT(\ell)\cdot\MET\left[1-\cos\left( \Delta\phi(\ell,\MET)\right)\right]},
\end{displaymath}
is required to be larger than 30~\GeV. Also, a more stringent cut on the lepton $\pT$ is applied to events in which the lepton and leading jet, $j_1$, are 
back-to-back, 
\begin{displaymath}
  \pT(\ell) > \mathrm{max}\left(25,40 \left( 1 - \frac{\pi - |\Delta \phi(j_1,\mbox{$\ell$})|}{\pi-1} \right)\right)~\mbox{\GeV},
\end{displaymath}
where $\Delta \phi(j_1,\ell)$ is the difference in azimuthal angle between the lepton momentum and the leading jet. 

The $W$ boson coming from the decay of the top quark can be reconstructed from the momenta of the lepton and the neutrino by using four-momentum conservation. The 
missing transverse momentum vector, with magnitude \MET, is used to represent the transverse component of the neutrino momentum and the longitudinal component 
$p_{\nu}^z$ is chosen such that the resulting $W$ boson is on its mass shell. A quadratic expression is found for $p_{\nu}^z$. If there are two possible real values, 
the value closer to zero is taken. If the solutions are complex, the assumption that the neutrino is the only contributor to the \MET\ is not valid. Therefore, the 
\MET\ is rescaled, preserving its direction, in order to have physical (real) solutions for $p_{\nu}^z$. If two solutions for \MET\ are found, the one resulting in 
the smaller value of $|p_{\nu}^z|$ is taken. The top-quark momentum is then reconstructed from this $W$ boson momentum and the momentum of the $b$-tagged jet. 
Finally, the momenta of the $W$ boson and spectator jet are boosted into the top-quark rest frame to obtain \Wmom\ and \Smom, used to generate the coordinate system 
in \Figure{fig:newaxes}, and the lepton is boosted into the $W$ boson rest frame to obtain \Lmom.

In addition to this basic event selection, further discrimination between \tch\ single top-quark events and background events is achieved by applying additional 
criteria: 
\begin{itemize}
  \item The pseudorapidity of the untagged jet must satisfy $|\eta|>2$, since the spectator jet tends to be forward in the \tch\ signature.
  \item The sum of the $\pT$ of all final-state objects, $H_{\text{T}}$, must be larger than 210~\GeV, since the $H_{\text{T}}$ distributions of the 
        backgrounds have their peaks just below this value.
  \item The mass of the top quark reconstructed from its decay products is required to be within 150--190~\GeV, to reject background events 
        from processes not involving top quarks.
  \item The distance in $\eta$ between the untagged jet and the $b$-tagged jet must be larger than 1, to further reduce \ttbar\ contributions.
\end{itemize}
These criteria and the basic event selection together define the signal region of the analysis.

Two control regions are defined, enhanced in each of the two dominant backgrounds, \ttbar\ and $W$+jets, to validate the modelling of the backgrounds by simulated 
events. In the control region where the \ttbar\ background is enriched, events with exactly four jets, exactly one of which is $b$-tagged, passing the basic event 
selection together with the top-quark mass and $H_{\text{T}}$ requirements are selected. In the control region where $W$+jets backgrounds are enriched, events with 
exactly two jets passing the basic event selection and an inversion of the top-quark mass criterion are required.

\section{Background estimation and event yields}
\label{sec:bkg_est}

The largest background contributions to single top-quark \tch\ production in the signal region arise from \ttbar\ production and from $W$ boson production in 
association with jets ($W$+jets). Events containing \ttbar\ production are difficult to distinguish from single top-quark events, as they also have real top quarks 
in the final state. Also, the production of a $W$ boson with two jets, where one is identified as containing a $b$-hadron, can mimic the \tch\ final state. \QCD\ 
production via the strong interaction can enter the signal region as well, and data-driven techniques are required to accurately model it, as explained at the end 
of this section. Other backgrounds originate from \Wtch\ and \sch\ single top-quark, diboson ($WW$, $WZ$, and $ZZ$) and $Z$+jets production. Most of these 
background models are derived directly from MC simulation, but specialised procedures are implemented for $W$+jets and \qcd\ production.

The \ttbar\ cross-section is calculated at NNLO in QCD including resummation of NNLL soft-gluon terms with Top++~2.0~\cite{Beneke:2011mq, Cacciari:2011hy, 
Baernreuther:2012ws, Czakon:2012zr, Czakon:2012pz, Czakon:2013goa, Czakon20142930} and is found to be $177^{+10}_{-11}$~pb. The uncertainties due to the PDFs at 
68\%~C.L. and \alphas\ are calculated following the PDF4LHC~\cite{Botje:2011sn} prescription for the \textsc{MSTW2008NNLO}, \textsc{CT10}, and 
\textsc{NNPDF2.3}~\cite{Ball:2010de} error PDF sets, and are added in quadrature to the scale uncertainty to yield a total uncertainty of 6\%. Single top-quark 
production cross-sections for \textit{Wt} and \sch\ production are calculated at approximate NNLO and are found to be 15.7$\pm 1.2$~pb~\cite{Kidonakis:2010ux} and 
4.63$^{+0.20}_{-0.18}$~pb~\cite{Kidonakis:2010tc}, respectively. The uncertainties correspond to the sum in quadrature of the uncertainty derived from the 
\textsc{MSTW2008NNLO} error PDF sets at 90\%~C.L. and the scale uncertainties, yielding a final uncertainty of about 8\% for \Wtch\ and about 4\% for \sch\ 
production. The $Z$+jets inclusive production cross-section is estimated with NNLO precision using the FEWZ program~\cite{Anastasiou:2003ds}. Diboson events ($WW$, 
$WZ$, and $ZZ$) are normalised to the NLO cross-section prediction calculated with MCFM~\cite{Campbell:2010ff}. The uncertainty on the combined $Z$+jets and diboson 
background is estimated to be 60\% as in Ref.~\cite{Aad:2014fwa}.

For the production of a $W$ boson in association with jets, the shapes of the relevant kinematic distributions are predicted from the \alpgen\ sample. They are 
initially normalised to make the inclusive $W$ cross-section correspond to the NNLO prediction using the FEWZ program~\cite{Anastasiou:2003ds}, with the same 
scaling factor being applied to the \alpgen\ prediction for the $W$+$b\bar{b}$, $W$+$c\bar{c}$, and $W$+light-jets samples. The \alpgen\ prediction for the $W$+$c$ 
process is scaled by a factor that is obtained from a study based on NLO calculations using MCFM~\cite{Campbell:2010ff}. Data-driven techniques are then used to 
estimate the flavour composition and the overall normalisation as described in Ref.~\cite{Aad:2012hg}, and an uncertainty of 18\% (15\%) for electron (muon) 
selections is applied. 

\QCD\ background events pass the signal region selection if a jet is mis-identified as an isolated lepton or if the event has a non-prompt lepton from the decay of 
a hadron that appears to be isolated. Since it is neither possible to simulate a sufficient number of those events nor to calculate the rate precisely, dedicated 
techniques are used to model \qcd\ events and to estimate their production rate, employing both collision data and simulated events. In the electron channel, 
mis-identified jets are the main source of \qcd\ background events. This is modelled by the jet--lepton method~\cite{Aad:2014fwa}, in which an electron-like jet is 
selected in \pythia\ di-jet MC with special requirements and redefined as a lepton. In the muon channel, a data-driven matrix method~\cite{Aad:2012qf,Aad:2014fwa} 
based on the muon impact parameter significance is used. An uncertainty of 50\% is assigned to the estimated yield based on comparisons of the rates obtained by 
using alternative methods~\cite{Aad:2014fwa}, i.e. the matrix method in the electron channel and the jet--lepton method in the muon channel, and using an 
alternative variable, i.e. $\mtletmiss$ instead of \MET. 

\begin{figure}[!hptb] 
  \centering
  \centerline{\makebox{
      \subfloat[]{\includegraphics[width=0.51\textwidth]{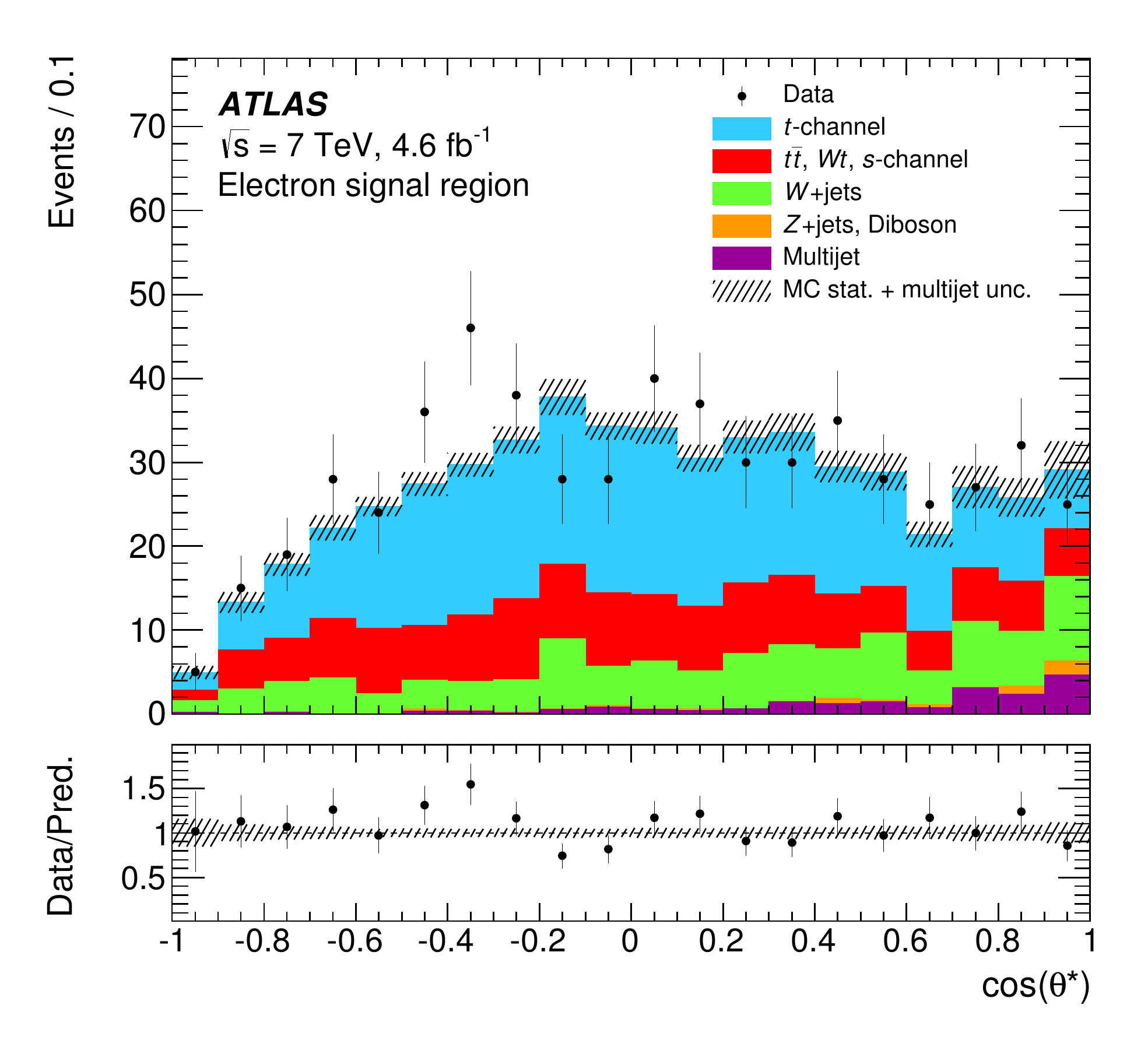}}
      \subfloat[]{\includegraphics[width=0.51\textwidth]{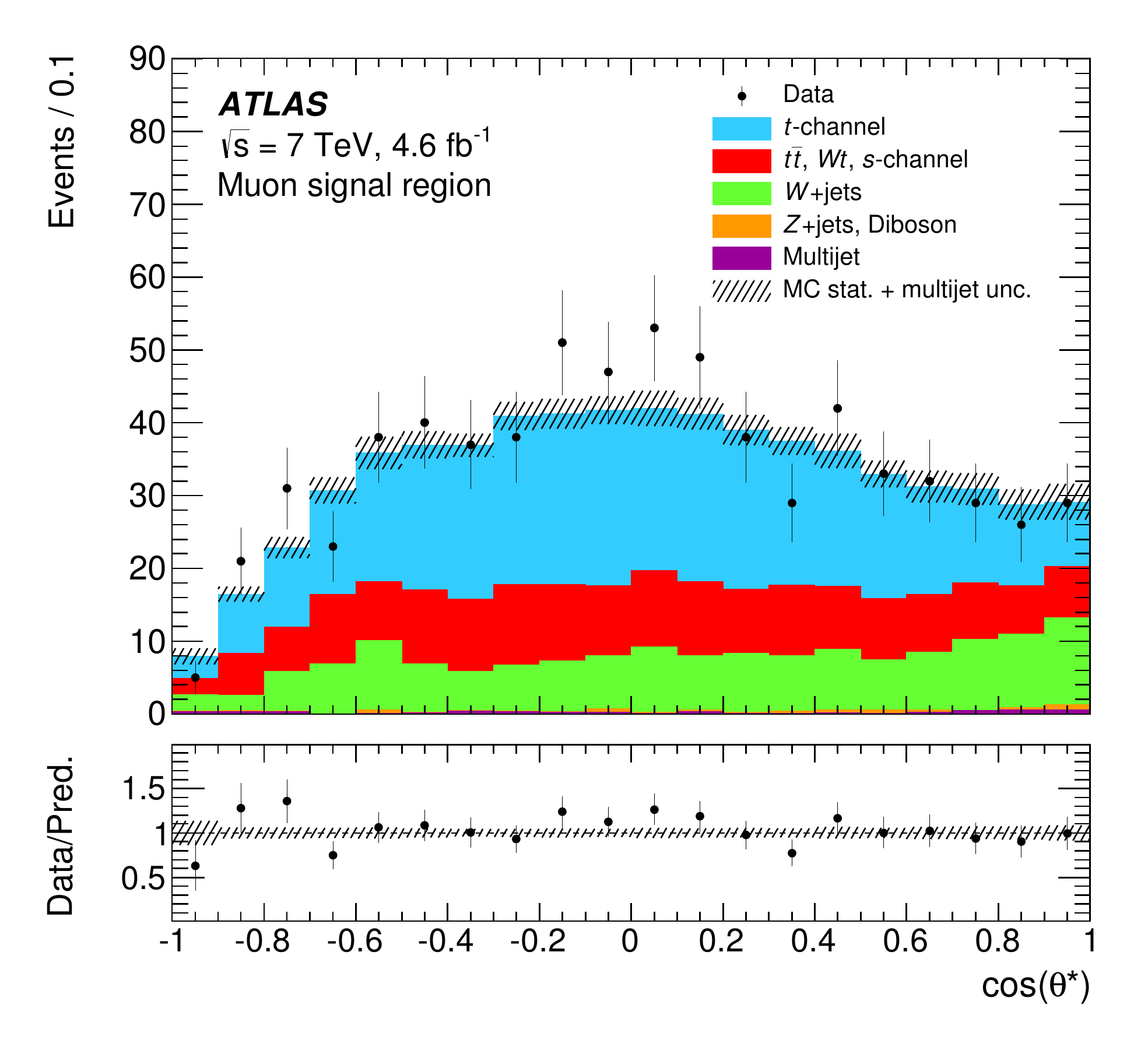}}
  }}
  \centerline{\makebox{
      \subfloat[]{\includegraphics[width=0.51\textwidth]{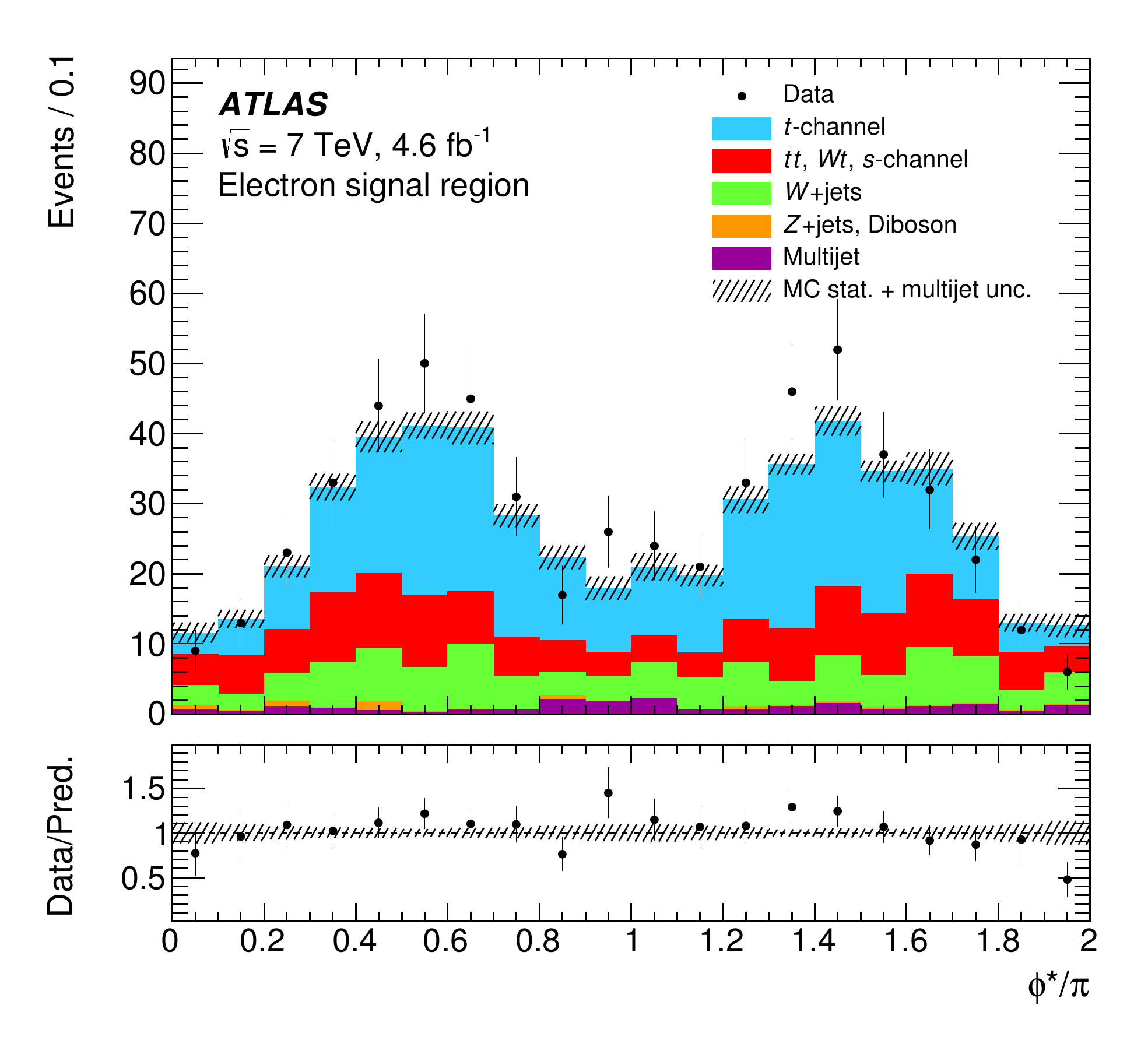}}
      \subfloat[]{\includegraphics[width=0.51\textwidth]{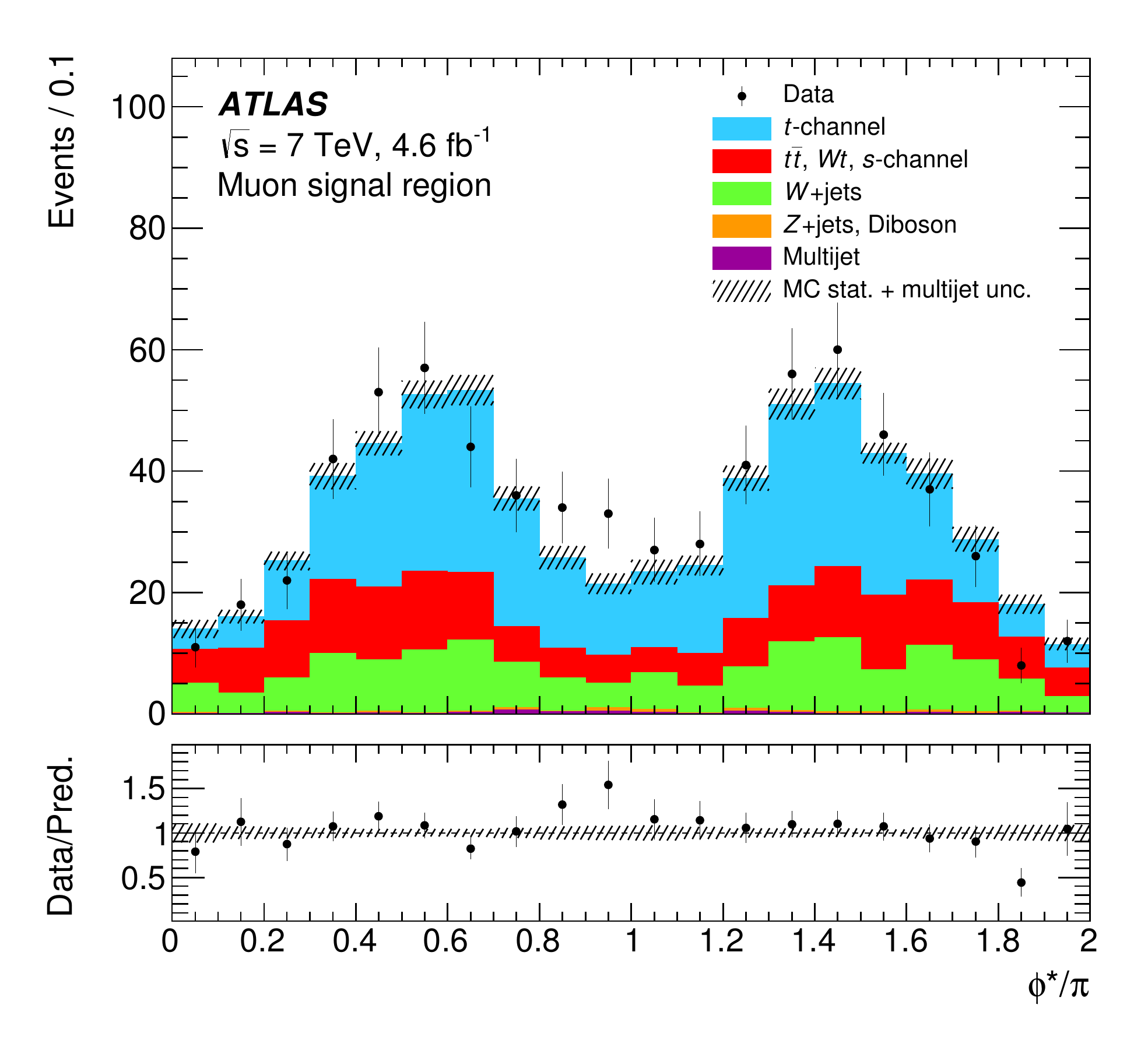}}
  }}
  \caption{ Angular distributions $\cosThetaS$ (upper row) and $\phiS$ (lower row) in the signal region for electrons (left) and muons (right), comparing data, 
            shown as the black points with statistical uncertainties, to SM signal and background predictions. The uncertainties shown on the prediction take into 
            account MC sample sizes and the 50\% systematic uncertainty on the normalisation of the \qcd\ background. The $W$+jets backgrounds are normalised to the 
            observed number of events in the control region, and the muon \qcd\ backgrounds are predicted from observed data. All other samples are scaled to their 
            theoretical predictions. The lower plots show the ratio of data to prediction in each bin. }
\label{fig:SR-AngDist_tch_sel_SM}
\end{figure}

\Table{tab:evtyield_both} provides the event yields for the electron
and muon channels after the event selection. The predictions are
normalised to \lumi\ and their uncertainties are derived taking into account the statistical uncertainties of the simulated samples. For the multijet background, the uncertainties on the prediction are the statistical uncertainties of the data and simulated samples of the muon and electron channel, respectively. \Figure{fig:SR-AngDist_tch_sel_SM} shows the 
angular distributions $\cosThetaS$ and $\phiS$ after the event selection for the electron and muon channels in data, compared to predictions using SM couplings. 
The uncertainties are derived taking into account
the statistical uncertainties of the data samples for data and simulated samples on the prediction and the 50\% systematic uncertainty on
the normalisation of the multijet background. Since the number of events in $W$+light-jets and \qcd\ samples are low, statistical fluctuations and events with large generator weights affect the sample shapes. Therefore, shape templates from events selected without the $b$-tagging requirement are used for these two backgrounds to smooth the simulated models. In these events, the hardest jet in the event is chosen to take the place of the $b$-tagged-like jet for the purposes of the event selection and reconstruction. Overall, good agreement between the observed and expected distributions is found.

\begin{table}[!htp]
  \begin{center}
    \caption{Event yields for the electron and muon channels in the
      signal region. Individual predictions are rounded to integers
      while ``Total expected'' corresponds to the rounding of the sum of full
      precision individual predictions. The uncertainties shown are statistical only. Uncertainties of less than 0.5 
             events appear as zero.}
    \vspace{2mm}
    \begin{tabular}{l r@{$\pm$}l | r@{$\pm$}l} 
\hline
\hline
 & \multicolumn{2}{c|}{Electrons} & \multicolumn{2}{c}{Muons} \T \B \\
\hline
\textit{t}-channel ~&~ 274~~&~2 ~&~ 336~~&~2 ~\\ 
\textit{s}-channel ~&~ 4~~&~0 ~&~ 5~~&~0 ~\\ 
\textit{Wt}-channel ~&~ 13~~&~1 ~&~ 16~~&~1 ~\\ 
\textit{t$\bar{t}$} ~&~ 119~~&~2 ~&~ 147~~&~3 ~\\ 
\textit{W}+heavy-jets ~&~ 92~~&~7 ~&~ 128~~&~8 ~\\ 
\textit{W}+light-jets ~&~ 13~~&~4 ~&~ 18~~&~3 ~\\ 
\textit{Z}+light-jets ~&~ 4~~&~1 ~&~ 3~~&~1 ~\\ 
Diboson ~&~ 1~~&~0 ~&~ 2~~&~0 ~\\ 
Multijet ~&~ 20~~&~7 ~&~ 6~~&~1 ~\\ 
\hline
Total expected ~&~ 538~~&~11 ~&~ 660~~&~10 \T \B \\ 
\hline
Data &\multicolumn{2}{c}{576}&\multicolumn{2}{c}{691} \T \B \\ 
\hline
\hline
\end{tabular}

    \label{tab:evtyield_both}
  \end{center}
\end{table}

\section{Analysis method}
\label{sec:analysis_method}

The model describing the \tch\ signal in \Section{sec:definition} is connected to the angles measured in reconstructed events via an analytic folding procedure. 
Models of the selection efficiency and detector resolution are derived from simulated \tch\ signal events, and a model of the reconstructed background is derived 
from the sum of the combined background processes described in \Section{sec:data_and_MC_samples}. Since the values of the parameters $\fu$ and $\delm$ depend on 
top-quark couplings, both the \tch\ signal and the \ttbar\ background are sensitive to their values. Only the angular distributions for the leptonic decay of the 
$W$ boson are measured, but the efficiency and resolution may also depend on other unmeasured distributions, such as those associated with the $\eta$ distribution 
of the top or spectator quark in the laboratory frame. The efficiency, resolution, and background models are constructed such that they account for any dependence 
on the values of $\fu$ and $\delm$, which is required to remove biases dependent on the values of these parameters (see Appendix \ref{app:param_dependence}). A 
probability density is derived for all events in the signal region, as a function of the reconstructed angles $\cosThetaS$ and $\phiS$, and conditional on the 
parameters. This density is then used to construct a likelihood, from which $\fu$ and $\delm$ are measured in data.

The signal model described by \Equation{eq:signal_model} is a series in spherical harmonics, only containing terms up to a maximum value of $l$, 
$l^\mathrm{sig}_\mathrm{max}=2$. Describing the efficiency and resolution models similarly allows the use of the orthogonality properties of spherical harmonics to 
construct analytically-folded distributions~\cite{Boudreau:2013yna}. An efficiency function, given by 
\begin{equation}
  \epsilon(\thetaST,\phiST;\ \params,P)=
      \sum\limits_{l^\prime,m^\prime}^{l^\mathrm{eff}_\mathrm{max}} e_{l^\prime,m^\prime}(\params,P)Y_{l^\prime}^{m^\prime}(\thetaST,\phiST), 
  \label{eq:efficiency_model}
\end{equation}
describes the probability that a \tch\ signal event with the given true angles $\thetaST$ and $\phiST$ will be reconstructed in the signal region. The series 
contains all allowed values of $l^\prime$ and $m^\prime$ up to a maximum spherical harmonic degree, $l^\mathrm{eff}_\mathrm{max}$. The selected signal density, 
$\rho_\mathrm{s}$, is then defined as the product of the efficiency function and the signal model, $\rho$, normalised to the overall rate,
\begin{equation}
  \begin{split}
    \rho_\mathrm{s}(\thetaST,\phiST;\ \params,P) 
      &= \frac{\epsilon(\thetaST,\phiST;\ \params,P)\rho(\thetaST,\phiST;\ \params,P)}
              {\int \epsilon(\thetaST,\phiST;\ \params,P)\rho(\thetaST,\phiST;\ \params,P) \mathrm{d}\Omega^*_\mathrm{T}}
       = \sum\limits_{L,M}c_{L,M}(\params,P) Y_L^M(\thetaST,\phiST), \\
      &\text{\ \ for\ \ \ } c_{L,M}(\params,P) =
          \frac{\sum\limits_{l,l^\prime,m,m^\prime} e_{l^\prime,m^\prime}(\params,P) a_{l,m}(\params,P) G_{l,l^\prime,L}^{m,m^\prime,M}}
               {\sum\limits_{l,m} {(-1)}^m e_{l,-m}(\params,P) a_{l,m}(\params,P)}.
  \end{split}
  \label{eq:selection_model}
\end{equation}
The Gaunt coefficients $G$ are defined in terms of Clebsch--Gordan coefficients $C_{\,l_1\, \,l_2\, \,L}^{m_1 m_2 M}$, given by,
\begin{displaymath}
  G_{l,l^\prime,L}^{m,m^\prime,M} = \sqrt{\frac{(2l+1)(2l^\prime+1)}{4\pi(2L+1)}} C_{l,l^\prime,L}^{0,0,0} C_{l,l^\prime,L}^{m,m^\prime,M}.
  \label{eq:gaunt_coefficients}
\end{displaymath}
The efficiency function is determined from a likelihood fit to simulated \tch\ events. A value is chosen for $l^\mathrm{eff}_\mathrm{max}$ by comparing the 
associated values of the corrected Akaike Information Criteria, AICc~\cite{Cavanaugh:1997201,Liddle:2007fy}, a likelihood-ratio test with an additional penalty 
term which increases with the number of parameters. From this test, $l^\mathrm{eff}_\mathrm{max}=6$ is chosen. The deviation in the results from the selection 
of efficiency models with $l^\mathrm{eff}_\mathrm{max}$ varied by $\pm2$ are included as systematic uncertainties. Other criteria, such as the Schwarz 
Criteria~\cite{Schwarz1978}, select values of  $l^\mathrm{eff}_\mathrm{max}$ within the range chosen for this uncertainty.

When reconstructing events in the signal region, the finite resolution of the detector results in a migration from the true angles to measured angles $\thetaS$ and 
$\phiS$. This migration is modelled by the resolution function,
\begin{equation}
  \mathcal{R}(\thetaS,\phiS|\thetaST,\phiST;\ \params,P) = \sum\limits_{\lambda,\mu}^{l^\mathrm{reco}_\mathrm{max}} 
             \sum\limits_{L^\prime,M^\prime}^{l^\mathrm{true}_\mathrm{max}} r_{\lambda,\mu,L^\prime,M^\prime}(\params,P) 
             Y_{\lambda}^{\mu}(\thetaS,\phiS) Y_{L^\prime}^{M^\prime}(\thetaST,\phiST).
  \label{eq:resolution_model}
\end{equation}
This series contains all terms with allowed values of $\lambda$, $\mu$, $L^\prime$, and $M^\prime$ up to the maximum degree parameters 
$l^\mathrm{true}_\mathrm{max}$, associated with the dependence on $\thetaST$ and $\phiST$, and $l^\mathrm{reco}_\mathrm{max}$, associated with the dependence on 
$\thetaS$ and $\phiS$. The reconstructed signal density, $\rho_\mathrm{r}$, is defined as the convolution of this function with the selected signal density, 
\begin{equation}
  \begin{split}
    \rho_\mathrm{r}(\thetaS,\phiS;\ \params,P) 
      &= \int\mathcal{R}(\thetaS,\phiS|\thetaST,\phiST;\ \params,P) \rho_s(\thetaST,\phiST;\ \params,P) \mathrm{d}\Omega^*_\mathrm{T} =
          \sum\limits_{\lambda,\mu}d_{\lambda,\mu}(\params,P) Y_{\lambda}^{\mu}(\theta^*,\phi^*), \\
      &\text{\ \ for\ \ \ } d_{\lambda,\mu}(\params,P) = \sum\limits_{L,M} (-1)^M c_{L,M}(\params,P) r_{\lambda,\mu,L,-M}(\params,P).
  \end{split}
  \label{eq:reconstruction_model}
\end{equation}
The resolution function is determined using a spherical Fourier technique~\cite{ref:OSDE}.
In principle, describing a very narrow resolution function could require large values of $l^\mathrm{true}_\mathrm{max}$ and $l^\mathrm{reco}_\mathrm{max}$.
In practice, the mathematics of angular-momentum addition guarantee that there are no terms in \Equation{eq:selection_model} with $L>l^\mathrm{sig}_\mathrm{max}+l^\mathrm{eff}_\mathrm{max}$.
Thus, $l^\mathrm{true}_\mathrm{max}=l^\mathrm{sig}_\mathrm{max}+l^\mathrm{eff}_\mathrm{max}$.
The parameter $l^\mathrm{reco}_\mathrm{max}$ is determined using the mean integrated squared 
error, $\mathrm{MISE} = \int \left(\rho(x)-\hat{\rho}(x)\right)^2 \mathrm{d}x$, where $\rho(x)$ is the true probability density and $\hat{\rho}(x)$ is a 
distribution estimating that density. Calculating the MISE with $\rho_\mathrm{r}$ yields a broad minimum for $l^\mathrm{reco}_\mathrm{max}>4$, and for this analysis 
$l^\mathrm{reco}_\mathrm{max}=6$ is chosen. The deviation in the results from the selection of resolution models with $l^\mathrm{reco}_\mathrm{max}$ varied by $\pm2$ are included as systematic uncertainties. 

This density of reconstructed \tch\ signal events is a series in spherical harmonics, with coefficients which are functions of the physics parameters describing the 
production ($P$) and decay ($\params$) of the top quark. A background model,
\begin{equation}
  \beta(\thetaS,\phiS;\ \params,P) = \sum\limits_{\lambda,\mu}^{l^\mathrm{bkg}_\mathrm{max}} b_{\lambda,\mu}(\params,P) Y_{\lambda}^{\mu}(\thetaS,\phiS), 
  \label{eq:background_model}
\end{equation}
is derived which is also a series in spherical harmonics, and is added to the reconstructed signal density via the signal fraction, $\fs$. This background function 
is determined from a likelihood fit to all simulated background events in the signal region, and contains all terms with allowed values of $\lambda$ and $\mu$ up to 
the maximum degree parameter $l^\mathrm{bkg}_\mathrm{max}$. This parameter is determined like the equivalent parameter in the efficiency function, using the AICc. 
The variations in the results from the selection of background models with $l^\mathrm{bkg}_\mathrm{max}\pm2$ are included as systematic uncertainties. 

The probability density of all events in the signal region, $\rho_\mathrm{t}$, is the sum of the reconstructed signal and background densities with signal fraction 
$\fs$,
\begin{equation}
  \begin{split}
    \rho_\mathrm{t}(\thetaS,\phiS;\ \params,P,\fs)    &= \sum\limits_{\lambda,\mu}^{\mathrm{max}\left(l^\mathrm{reco}_\mathrm{max},l^\mathrm{bkg}_\mathrm{max}\right)} 
                                                \mathcal{A}_{\lambda,\mu}(\params,P) Y_{\lambda}^{\mu}(\thetaS,\phiS) \text{\ for} \\
    \mathcal{A}_{\lambda,\mu}(\params,P,\fs) &= f_s d_{\lambda,\mu}(\params,P) + (1-f_s) b_{\lambda,\mu} \\
     &= f_s \frac{ \sum\limits_{l,l^\prime,L,m,m^\prime,M} (-1)^M e_{l^\prime,m^\prime} a_{l,m}(\params,P) G_{l,l^\prime,L}^{m,m^\prime,M} r_{\lambda,\mu,L,-M} } 
                 { \sum\limits_{l,m} (-1)^m e_{l,-m} a_{l,m}(\params,P) } + (1-f_s) b_{\lambda,\mu}.
  \end{split}
  \label{eq:measurement_model}
\end{equation}

\begin{figure}
  \centering
  \centerline{\makebox{
      \subfloat[]{\includegraphics[width=\textwidth]{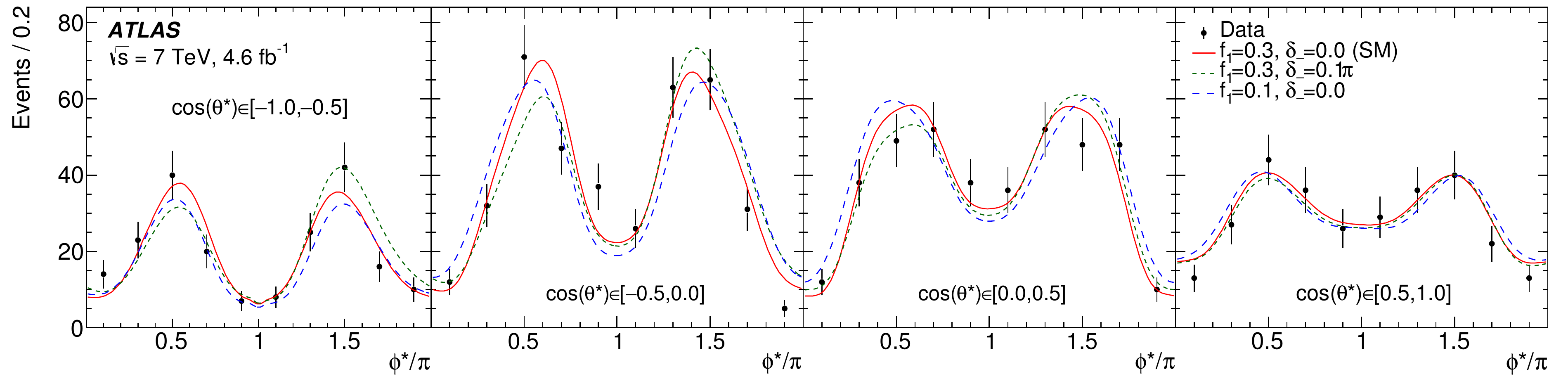}}
    }}
  \centerline{\makebox{
      \subfloat[]{\includegraphics[width=3.2in]{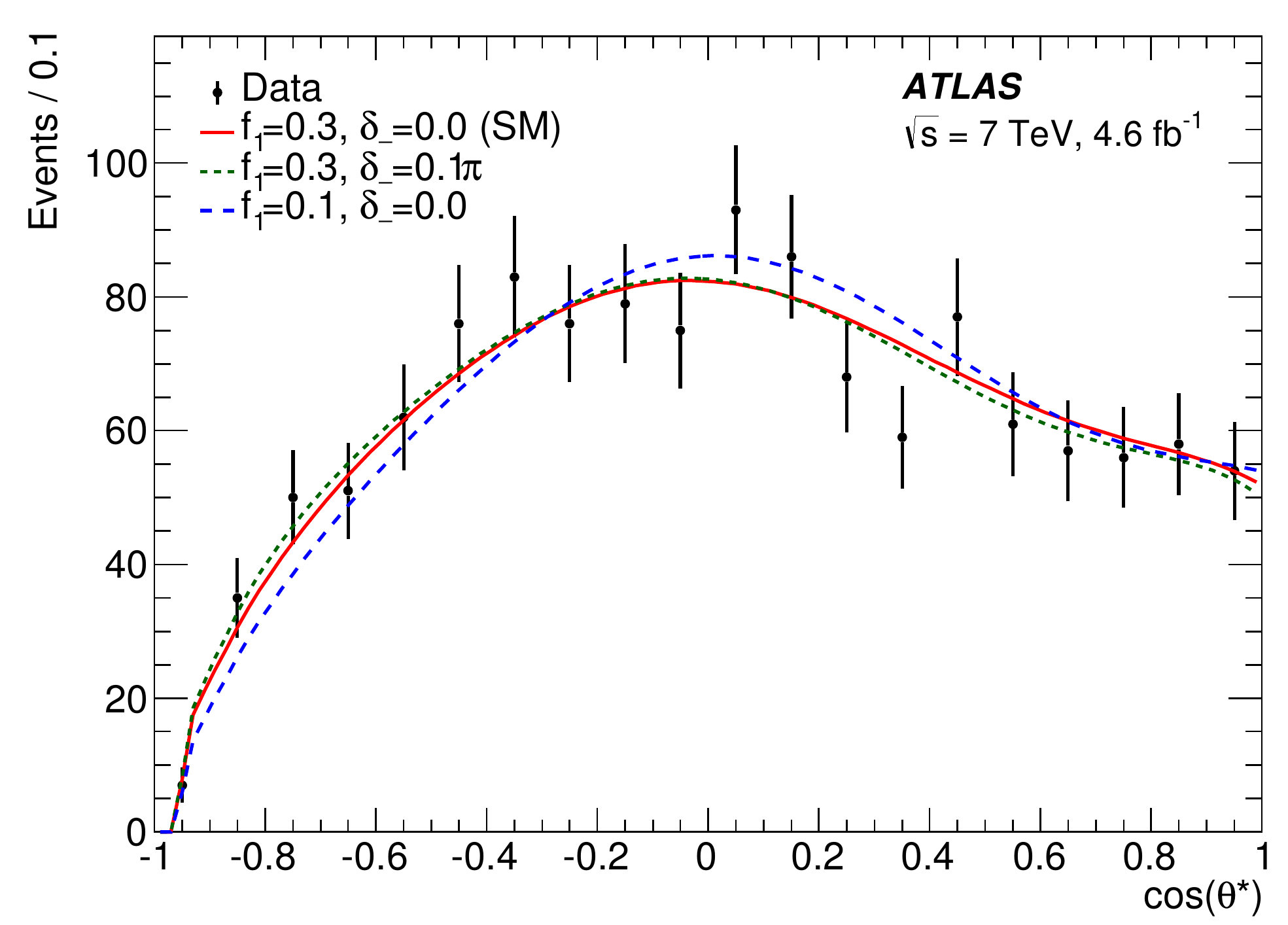}}
      \subfloat[]{\includegraphics[width=3.2in]{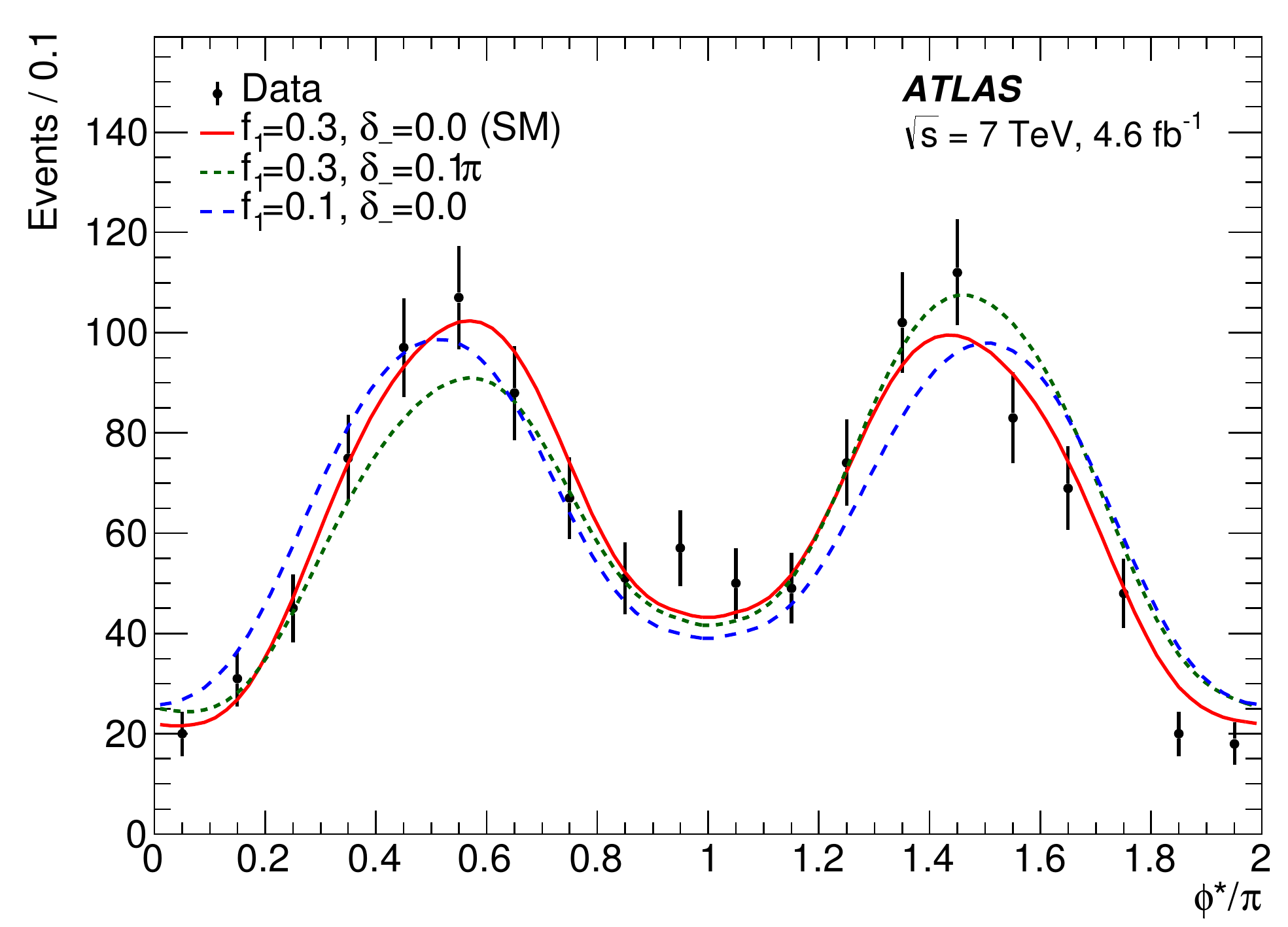}}
    }}
  \caption{ Projections into (a) $\phiS$ in bins of $\cosThetaS$, (b) $\cosThetaS$, and (c) $\phiS$ of the function described in \Equation{eq:measurement_model}, 
            for different values of $\fu$ and $\delm$. The black points shown are for the selected data events with statistical uncertainties. The curves shown 
            represent the model at the SM point $\fu=0.3$, $\delm=0$ (solid red), and two sets of beyond-the-SM values, $\fu=0.1$, $\delm=0$ (dashed blue) and 
            $\fu=0.3$, $\delm=0.1\pi$ (dotted green). }
\label{fig:finModelVariation}
\end{figure}

Figure \ref{fig:finModelVariation} shows the data summed over the electron and muon channels for the signal region. Three representative models are also shown: one 
for the SM expectations, one that is near the expected 95\%~C.L. sensitivity in $\fu$, and one near the expected sensitivity in $\delm$. These values for $\fu$ and 
$\delm$ are the same as those used in Figure
\ref{fig:genModelVariation}, but the curves now include the effects of
efficiency, resolution and background. 
The main differences between the two figures are due to the isolation requirements placed on the leptons. 
For $\cosThetaS=-1$, the lepton overlaps the $b$-tagged jet, while for $\phiS=\pi$, the lepton may overlap the spectator jet.
The region where the efficiency is less than 0.05\% ($\cosThetaS <-0.95$, and $-0.95<\cosThetaS <-0.85$ for $\pi-0.1<\phiS <\pi+0.1$), is excluded from further analysis.
For other regions, the efficiency varies, as a function of $\phiS$ and $\cosThetaS$, from 0.1\% up to
1.7\% for larger $\cosThetaS$ and away from the plane containing the spectator jet.
As the features of the angular
distributions are broad, the effects of resolution are small. The main
effect is to smooth out some of the effects of the acceptance and to
increase the number of events at large $\cosThetaS$ where the SM contribution is small.
As seen in Figure \ref{fig:SR-AngDist_tch_sel_SM}, the background is also somewhat larger for large $\cosThetaS$.

An unbinned likelihood function is constructed from the probability density in \Equation{eq:measurement_model} and from a set of events 
$\mathcal{D}=\left\{\thetaS_i,\phiS_i,w_i\right\}$, $i=1,2,...N$,
\begin{equation}
  \mathcal{L}(\params) = \prod\limits_{i=1}^{N} \exp\left({w_i \rho_\mathrm{t}(\theta^*_i,\phi^*_i|\ \params,P,\fs) }\right),
  \label{eq:measurement_likelihood}
\end{equation}
where $\thetaS_i$ and $\phiS_i $ can be taken either from data or MC simulation, and the event weights $w_i$ are 1 for measured data. The likelihood is evaluated 
on a grid with spacing 0.01 in $\fu$ and 0.01$\pi$ in $\delm$. The efficiency and resolution models at each point are derived from \protos\ \tch\ simulated events, 
and the background model uses \protos\ \ttbar\ simulated events. Interpreting these parameters as varying the coupling ratio $\gr/\vl$, the polarisation $P$ is 
varied simultaneously according to the parameterisation in Ref.~\cite{ref:ASPolPaper}. 

A central value is measured at the maximum value of $\mathcal{L}$ on this grid, determined from a Gaussian fit to the points at which $\mathcal{L}$ is evaluated. 
The 68\% and 95\% confidence limits are defined by the region where the likelihood ratio, $\mathcal{L}/\mathcal{L}_\mathrm{max}$, is larger than the value that 
would yield the same likelihood for a two-dimensional Gaussian distribution. They are refined to increase the accessible precision by interpolating between points 
on either side of the contours determined from these evaluated points. The statistical uncertainty is estimated from the symmetrised 68\%~C.L. interval of each 
parameter. 

This model is checked for closure both with MC samples, and by defining toy-MC samples based on \Equation{eq:measurement_model} with the same number of events as in 
data. These samples were generated at various points in $(\fu,\delm)$ space. In all cases they are found to reproduce the expected values of $\fu$ and $\delm$. These 
toy-MC samples were also used to derive pull distributions for $\fu$ and $\delm$ to check the statistical uncertainties returned by the fit.

\section{Sources of systematic uncertainty}
\label{sec:systematics}

Systematic uncertainties are evaluated in the full $(\fu,\delm)$ parameter space. Efficiency, resolution, and background models are determined from MC samples with 
a parameter varied by its uncertainty, or a subset where appropriate. A likelihood is constructed from the resulting model, using events generated with nominal 
values of the varied parameters. The difference between the central values estimated at the nominal value of a parameter and at the value varied by its uncertainty, 
or half the difference between central values estimated with the parameter varied up and down by its uncertainty, is used to construct a covariance matrix for each 
source of systematic uncertainty. The total covariance matrix for the systematic uncertainties and its correlation matrix are found from the sum of the covariance 
matrices determined for individual uncertainties. 
 
\subsection{Object modelling}
\label{sec:syst_objects}
The uncertainties in the reconstruction of jets, leptons, and \MET\ are propagated to the analysis, following the same procedures as described in 
Ref.~\cite{Aad:2014fwa}. The main source of uncertainty from these physics objects is the jet energy scale (JES)~\cite{JES2011}, which is the largest systematic 
uncertainty on the measurement of $\fu$. To estimate the impact of the JES uncertainty on the result, the jet energy is scaled up and down by its 
uncertainty~\cite{JES2011}, which ranges from 2.5\% in the central region with high-\pT\ jets to 14\% in the far forward region with low-\pT jets. Uncertainties 
are also estimated for jet energy resolution and reconstruction efficiency; the impact of varying the jet-vertex fraction requirement; \MET\ reconstruction and the 
effect of pile-up collisions on the \MET; $b$-tagging efficiency and mistagging rate; lepton trigger, identification, and reconstruction efficiencies; and lepton 
momentum, energy scale, and resolution. 

\subsection{MC generators and PDFs}
\label{sec:syst_generators}
Multiple MC event generators are used to model the \tch\ and \ttbar\ processes in this analysis, and the differences between these generators are included as 
systematic uncertainties. Comparing the \acermc\ generator used for \tch\ events and the \powhegbox\ generator used for \ttbar\ events to the \protos\ generator 
used for both the \tch\ single top-quark events and \ttbar\ events yields the largest uncertainty in $\delm$. Additional \tch\ comparisons are performed between 
the \acermc\ and \powhegbox\ generators and between \acermc\ events with showering by \pythia\ and \herwig. The renormalisation and factorisation scales in the 
\powhegbox\ \tch\ sample are also varied independently by factors of 0.5 and 2.0. Additional \ttbar\ comparisons are performed between the \powhegbox\ generator 
and the \mcanlo~\cite{SAMPLES-MCANLO}\ generator with showering by \herwig, and between \powhegbox\ events with showering by \pythia\ and \herwig. The variations 
between PDF sets and within individual sets are used to estimate a systematic uncertainty following the PDF4LHC prescription. 

\subsection{Signal and background normalisation}
\label{sec:syst_background}
Cross-sections for background processes given in \Section{sec:bkg_est} are varied up and down by their uncertainties. The \qcd\ normalisation is varied by 50\% as 
discussed in \Section{sec:bkg_est}, and the remaining backgrounds are varied simultaneously to produce a conservative estimate. The impact of object modelling and 
other uncertainties on the $W$+jets background normalisation are considered in parallel with the variations made in \tch\ and \ttbar\ samples. A separate shape 
uncertainty is assigned to the $W$+jets samples by varying the matching and factorisation scales in the \alpgen\ generator. The integrated luminosity is varied up 
and down by its uncertainty, $\pm1.8$\%, derived as detailed in Ref.~\cite{ref:AtlasLumiPaper2013}.

\subsection{Detector correction and background parameterisation}
\label{sec:syst_model}
An uncertainty due to the limited size of the MC samples used to estimate the efficiency and resolution models is estimated from the statistical uncertainties 
derived from a measurement of $\fu$  and $\delm$ in the \tch\ signal MC sample alone. The background statistical uncertainty is estimated by varying the background 
model according to the eigenvectors of its covariance matrix. The background statistical uncertainty dominates the total ``MC statistics'' uncertainty listed in 
\Table{tab:systuncs}. Its significance reflects the small size of some background samples in the signal region, and the resulting disparate values of the sample 
weights. The effect of varying the cutoff degree $l_\mathrm{max}$ in the determination of these models is also estimated, and is found to be small compared to MC 
statistical uncertainty.

\subsection{Uncertainty combination}
\label{sec:syst_combined}
\Table{tab:systuncs} shows the contribution of each source of uncertainty to the measurement of the parameters $\fu$ and $\delm$ and their correlation, 
$\rho(\fu,\delm)$. The total systematic uncertainty and correlation is obtained from the sum of the covariance matrices determined for each source. It is combined 
with the covariance matrix of the statistical uncertainty via the following method. At each point $\vec{\alpha}_i$ in the $(\fu,\delm)$ space, a multivariate normal 
distribution $\mathcal{N}_i$ is constructed with the covariance matrix representing the systematic uncertainties, $\mathbf{\Sigma}^\mathrm{syst}$, and the mean, 
$\vec{\alpha}_i$. The resulting distribution is evaluated at a point $\vec{\alpha}_j$ and multiplied by the likelihood at this point, $\mathcal{L}^\mathrm{stat}_j$. 
The maximum modified likelihood value, over all possible points $\vec{\alpha}_j$, is kept, and the resulting broadened likelihood distribution 
$\mathcal{L}^\mathrm{stat+syst}_i$ is used to represent the measurement with both statistical and systematic variation incorporated:
\begin{equation}
  \mathcal{L}^\mathrm{stat+syst}_i = \mathrm{max}_j\left\{ \mathcal{N}_i(\vec{\alpha_j} ;\vec{\alpha}_i,\mathbf{\Sigma}^\mathrm{syst})\cdot\mathcal{L}^\mathrm{stat}_j\right\}.
  \label{eq:syst_comb}
\end{equation}
The resulting confidence limits are taken as the full uncertainty in the measurement. 

\begin{table}[ht!]
  \centering
  \caption{ Sources of systematic uncertainty on the measurement of $\fu$ and $\delm$ using \acermc\ \tch\ single top-quark simulated events and backgrounds 
            estimated from both MC simulation and data, including \powhegbox\ \ttbar\ simulation. Individual sources are evaluated separately for shifts up and 
            down, and symmetrised uncertainties $\sigma(\fu)$, $\sigma(\delm)$ and correlation coefficients $\rho(\fu,\delm)$ are given. }
  \vspace{2mm}
  \begin{tabular}{l|r|r|r} \hline \hline
  Source & \multicolumn{1}{c|}{$\sigma(f_1)$} & \multicolumn{1}{c|}{$\sigma(\delta_-)/\pi$} & \multicolumn{1}{c}{$\rho(f_1,\delta_-)$} \T \B \\ \hline \hline
  Data statistics & \fustat & \delmpistat & \corrparamstat \T \B \\ \hline \hline

  Jets &  0.03  & 0.015 & \multicolumn{1}{r}{0.39} \T \\
  $b$-tagging & $<0.01$ & $<0.001$ & \multicolumn{1}{r}{$-$0.70} \\
  Leptons & 0.02 & 0.007 & \multicolumn{1}{r}{0.39} \\
  \MET & 0.01 & 0.004 & \multicolumn{1}{r}{$-0.27$} \B\\ \hline

  Generator & 0.02 & 0.017 & \multicolumn{1}{r}{0.40}  \T \\
  Parton shower & 0.02 & 0.001 & \multicolumn{1}{r}{0.98}  \\
  PDF variations & 0.01 & 0.009 & \multicolumn{1}{r}{0.23} \B \\ \hline

  Cross-sections & $<0.01$ & $<0.001$ & \multicolumn{1}{r}{1.00}  \T \\
  $W$+jets shape & $<0.01$ & 0.001 & \multicolumn{1}{r}{$-0.59$} \\
  Multijet normalisation & $<0.01$ & 0.002 & \multicolumn{1}{r}{$-1.00$}  \\ 
  Luminosity & $<0.01$ & $<0.001$ & \multicolumn{1}{r}{$-1.00$} \B \\ \hline

  Model $l_\mathrm{max}$ variation & 0.01 & 0.001 & \multicolumn{1}{r}{$-0.70$}  \T \\
  MC statistics & 0.02 & 0.011 & \multicolumn{1}{r}{0.14} \B \\ \hline \hline

  Combined systematic & 0.05 & 0.028 & \multicolumn{1}{r}{0.27} \T \B \\ \hline \hline
  Total & \fufull & \delmpifull & \multicolumn{1}{r}{\corrparamfull} \T \T \B \B \\ \hline \hline
\end{tabular}

\label{tab:systuncs}
\end{table}

\section{Results}
\label{sec:result}

\begin{figure}[!htb]
  \centering
  \centerline{\makebox{
    \subfloat[]{\includegraphics[width=3.2in]{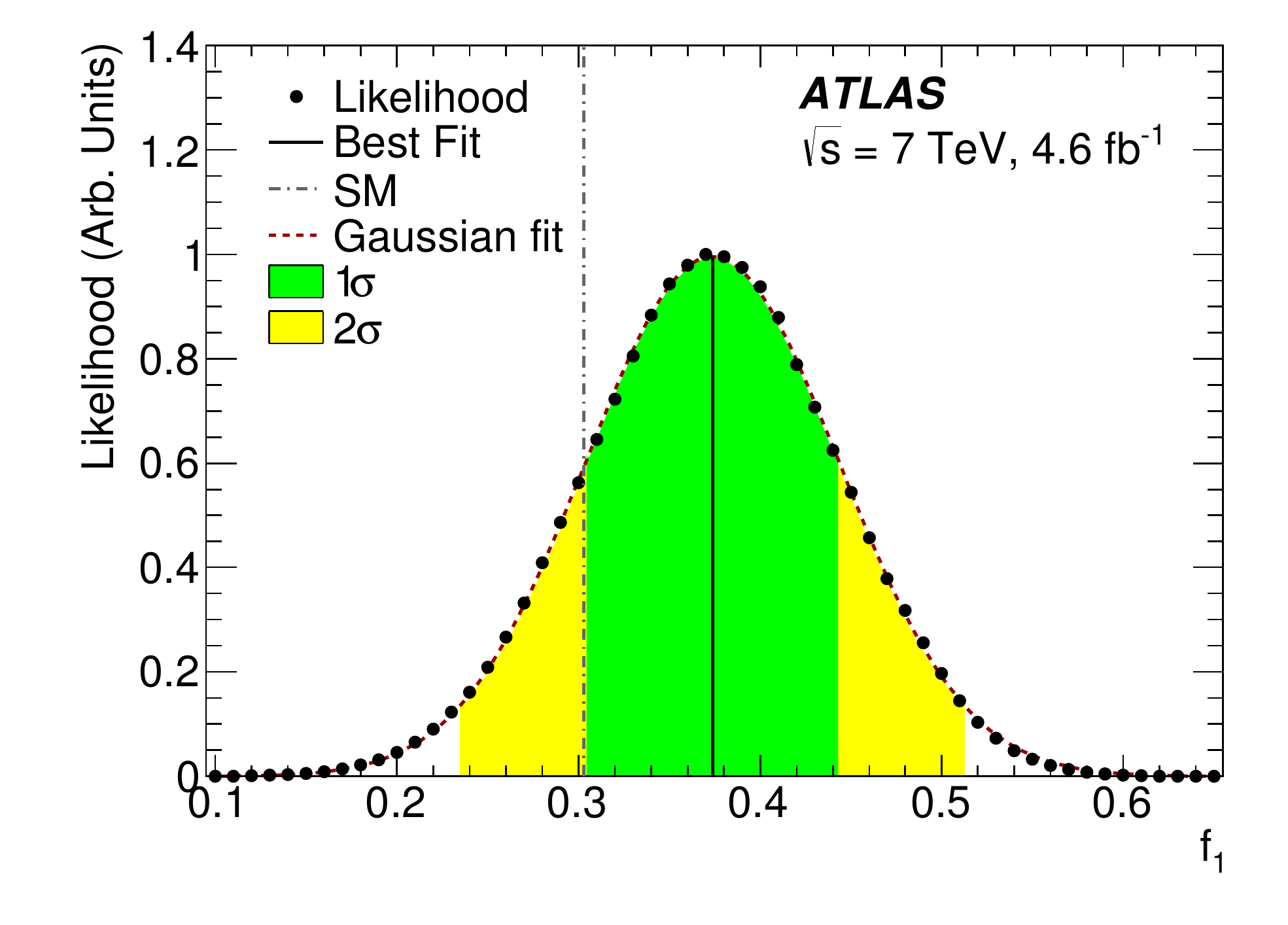}}
    \subfloat[]{\includegraphics[width=3.2in]{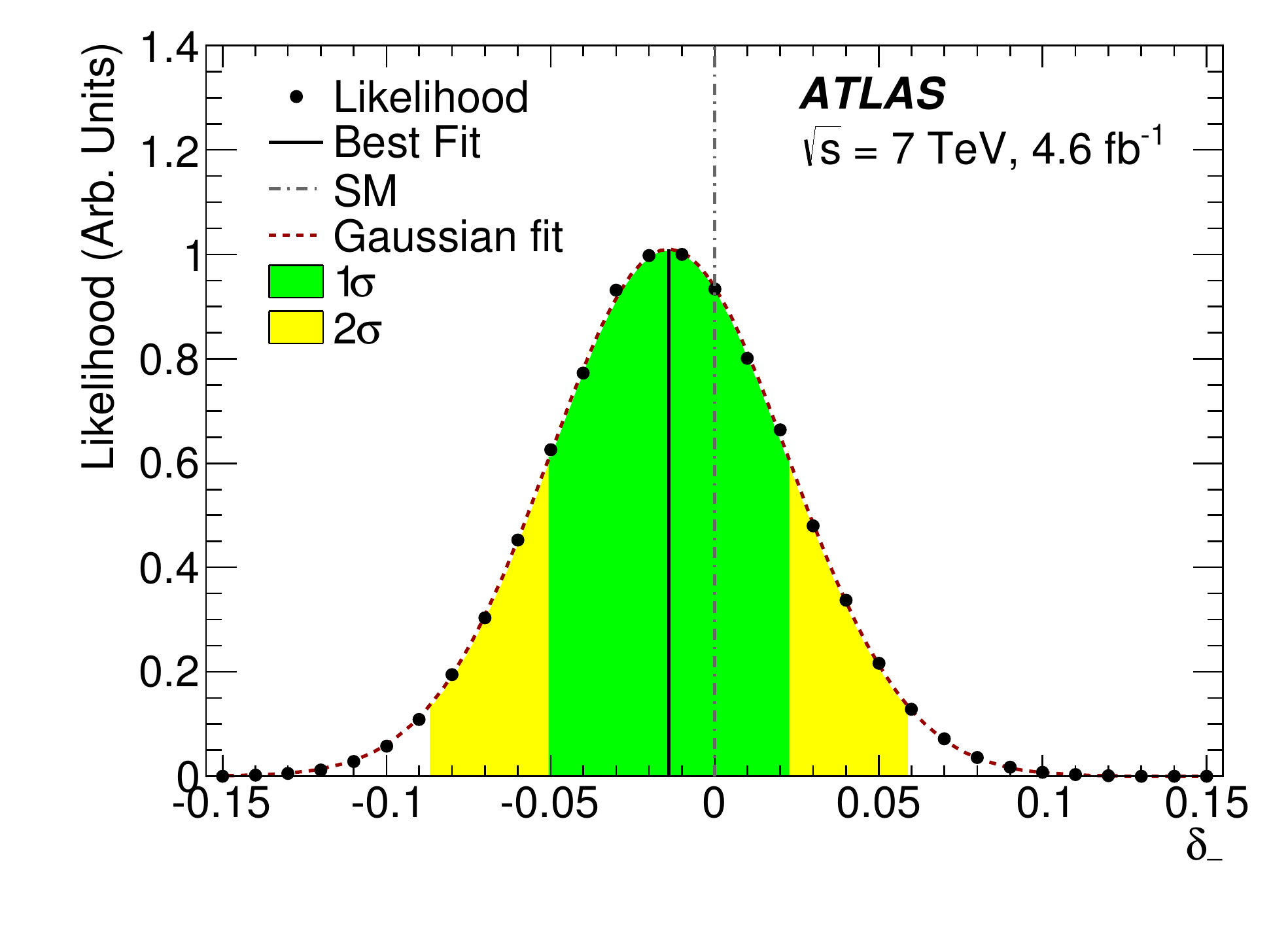}}
  }}
  \centerline{\makebox{
    \subfloat[]{\includegraphics[width=3.2in]{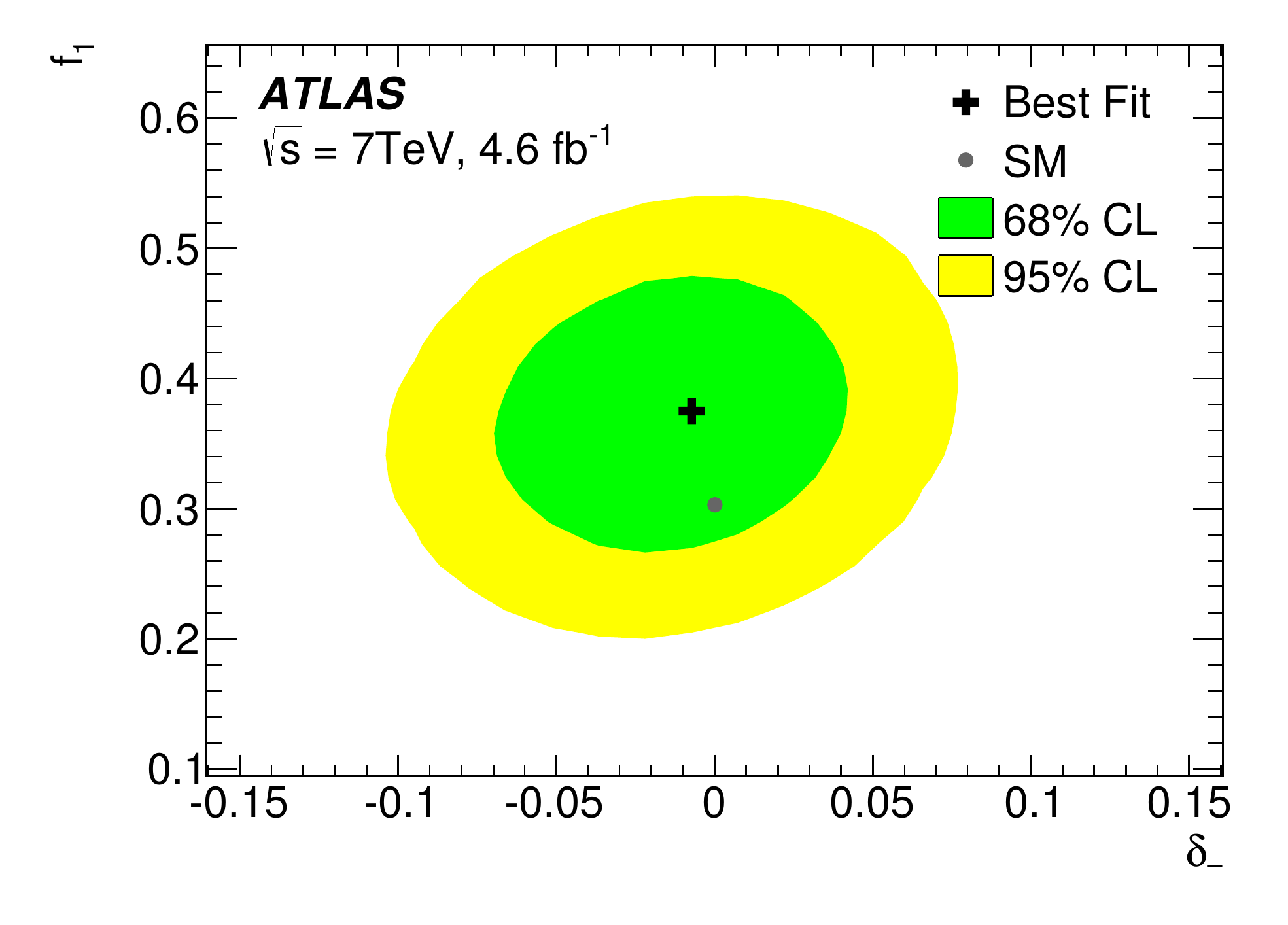}}
    \subfloat[]{\includegraphics[width=3.2in]{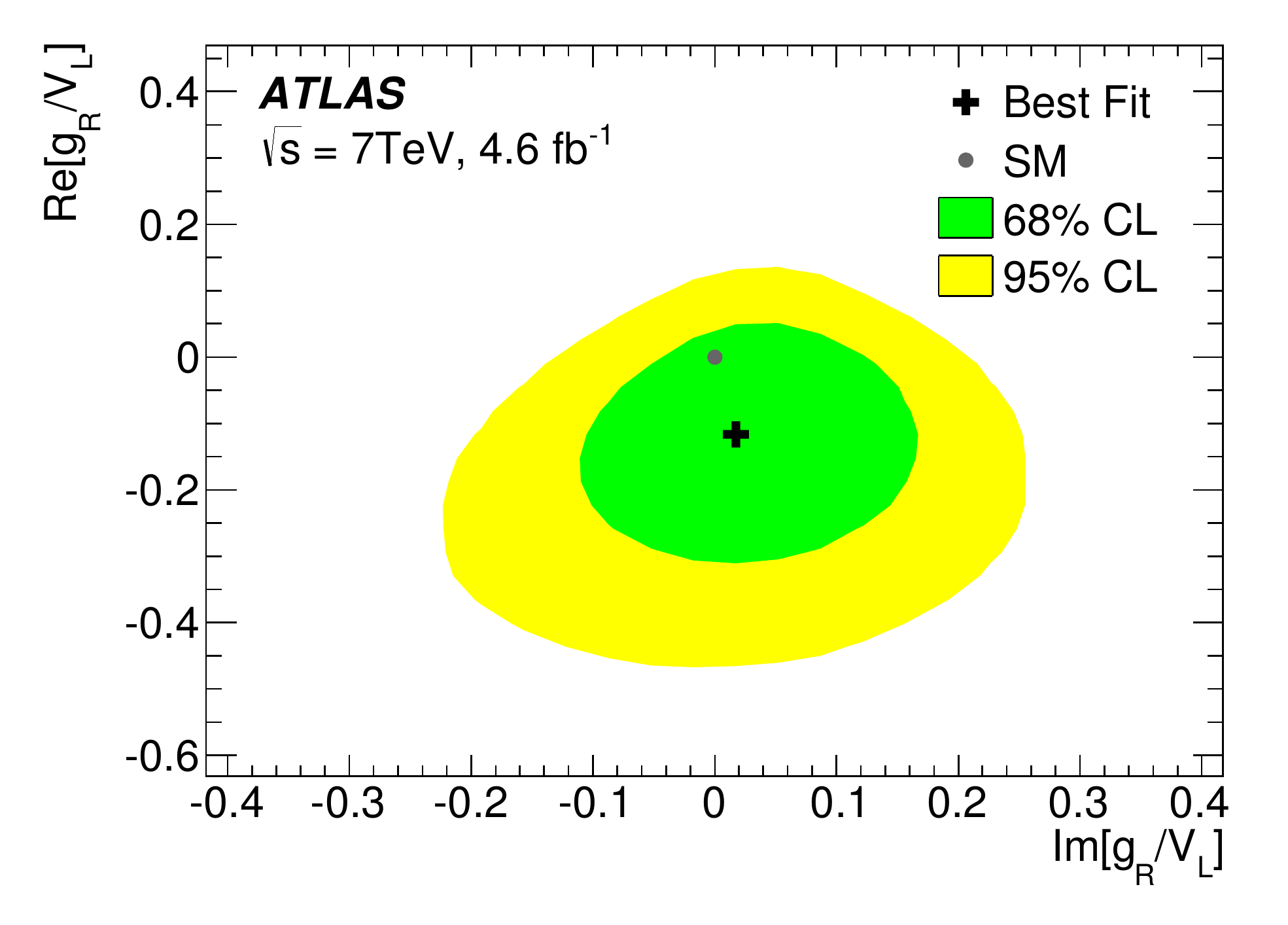}}
  }}
  \caption{ Projections of the likelihood function constructed from the signal region probability density \Equation{eq:measurement_model} and data events into 
            (a) $\fu$, (b) $\delm$, (c) $\fu$ vs. $\delm$, and (d) $\regrvl$ vs. $\imgrvl$, with systematic uncertainties incorporated. The black points indicate 
            the largest evaluated likelihood in each bin of the projected variable. Gaussian fits to the one-dimensional projections were performed, displayed as 
            the red curve. Regions shown in green and yellow represent the 68\% and 95\% confidence level regions, respectively. A black line or cross indicates 
            the observed value, and the grey line or point indicates the SM expectation. }
\label{fig:measCVFull}
\end{figure}

The result for $(\fu,\delm)$ and the coupling ratios $(\regrvl,\imgrvl)$ is shown in \Figure{fig:measCVFull}. The 68\% contour represents the total 
uncertainty on the measurement.

The parameters $\fu$ and $\delm$ and their uncertainties are measured to be 
\begin{equation}
  \begin{aligned}
    &\fu                       &=& \quad \cvfufullunc,  \\ 
    &\delm                     &=& \quad \cvdelmfullunc. 
  \end{aligned}
  \label{eq:result_parameters}
\end{equation}
The correlation in the measurement of these parameters is $\rho\left(\fu,\delm\right)=\corrparamfull$. The results are compatible with the SM expectations at LO, 
derived from expressions in Refs.~\cite{Fischer:2001gp,AguilarSaavedra:2010nx} with $m_t=172.5$~\GeV, $\mW=80.399$~\GeV, and $m_b=4.95$~\GeV: $\fu=0.304$ and 
$\delm=0$.

The dependence of the parameters $\fu$ and $\delm$ on the top-quark mass is evaluated using \tch\ and \ttbar\ simulation samples with a range of different top-quark 
masses. A linear dependence is found, resulting from changes in acceptance at different masses, with a slope of $-0.019$~\GeV$^{-1}$ for $\fu$ and a negligible 
slope for $\delm$. The uncertainty due to the top-quark mass dependence is not included in the total systematic uncertainty since it has no significant impact on 
the results.

The propagation of the uncertainties to the $(\regrvl,\imgrvl)$ space gives 
\begin{equation}
  \begin{aligned}
    &\regrvl                         &=& \quad \cvregrvlfullunc, \\
    &\imgrvl                         &=& \quad \cvimgrvlfullunc.
  \end{aligned}
  \label{eq:result_couplings}
\end{equation}

The correlation in the measurement of these coupling ratios is $\rho\left(\regrvl,\imgrvl\right)=\corrcoupfull$. The effect on the propagation due to the current 
uncertainty in the top-quark, $W$ boson and $b$-quark masses~\cite{Agashe:2014kda} is $<0.01$ in $\regrvl$, and $<0.0001$ in $\imgrvl$.
\par

Limits are placed simultaneously on the possible complex values of the ratio of the anomalous couplings \gr\ and \vl\ at 95\%~C.L., 
\begin{equation}
  \regrvl\in\regrvllimits \quad \mathrm{and} \quad \imgrvl\in\imgrvllimits. 
  \label{eq:result_limits}
\end{equation}

The best constraints on $\regr$ come from $W$ boson helicity fractions in top-quark decays, with $\regr$ of $[-0.08,0.04]$ and $[-0.08,0.07]$, both at 95\%~C.L., 
from ATLAS~\cite{Aad:2012ky} and from CMS~\cite{ref:CMSWhellplusJets7TeV}, respectively. However, these limits use the measured single top-quark production 
cross-section~\cite{Aad:2014fwa,Chatrchyan:2012ep} along with the assumption that $\vl=1$ and $\imgr=0$. Without these assumptions no value within the range 
$0.0 \lesssim \regrvl \lesssim 0.8$ can be excluded. The limits presented in this paper remove these assumptions and extend the knowledge of $\gr$ to the whole 
complex plane by simultaneously measuring information about $\regrvl$ and $\imgrvl$; the latter is previously unmeasured.

\FloatBarrier

\section{Conclusion}
\label{sec:conclusion}

The analysis described in this publication is a two-dimensional measurement of anomalous couplings in the \Wtb\ vertex in \tch\ single top-quarks events, performed 
on \lumi\ of \textit{pp} collisions at a centre-of-mass energy of \lhcenergy\ collected by the ATLAS detector at the LHC. An analytic folding model and likelihood 
maximisation techniques are used to extract values of the parameters \fu\ and \delm\ in a parameterisation of the decay of the top quark, which translate to limits 
on the couplings \regrvl\ and \imgrvl\ with $\vr=\gl=0$. The coupling parameterisation in terms of the spherical angles $\thetaS$ and $\phiS$ defines the angular 
distribution of the \tch\ signal process. Efficiency and resolution functions are fit in this space, which are then folded into the signal model describing the 
underlying physics analytically. A background function is fit and added to the signal model. The full model is used to construct a likelihood and its 
characterisation provides estimators of the best-fit central value, uncertainties, and correlations between \fu\ and \delm, or \regrvl\ and \imgrvl. The result is 
combined with estimated sources of systematic uncertainty to give the measurement of the parameters, $\fu=\cvfufullunc$, $\delm=\cvdelmfullunc$, and 
$\rho(\fu,\delm)=\corrparamfull$, or $\regrvl=\cvregrvlfullunc$, $\imgrvl=\cvimgrvlfullunc$, and $\rho(\regrvl,\imgrvl)=\corrcoupfull$ in terms of the couplings. 
The measurement sets limits at the 95\% confidence level on $\regrvl\in\regrvllimits$  and $\imgrvl\in\imgrvllimits$. The limits on \regrvl\ are complementary to 
limits from $B$-meson decay and $W$ helicity measurements, and contain previously unmeasured information about \imgrvl. The measured values are in agreement with 
SM expectations.

\section*{Acknowledgements}

We thank CERN for the very successful operation of the LHC, as well as the
support staff from our institutions without whom ATLAS could not be
operated efficiently.

We acknowledge the support of ANPCyT, Argentina; YerPhI, Armenia; ARC, Australia; BMWFW and FWF, Austria; ANAS, Azerbaijan; SSTC, Belarus; CNPq and FAPESP, Brazil; NSERC, NRC and CFI, Canada; CERN; CONICYT, Chile; CAS, MOST and NSFC, China; COLCIENCIAS, Colombia; MSMT CR, MPO CR and VSC CR, Czech Republic; DNRF, DNSRC and Lundbeck Foundation, Denmark; IN2P3-CNRS, CEA-DSM/IRFU, France; GNSF, Georgia; BMBF, HGF, and MPG, Germany; GSRT, Greece; RGC, Hong Kong SAR, China; ISF, I-CORE and Benoziyo Center, Israel; INFN, Italy; MEXT and JSPS, Japan; CNRST, Morocco; FOM and NWO, Netherlands; RCN, Norway; MNiSW and NCN, Poland; FCT, Portugal; MNE/IFA, Romania; MES of Russia and NRC KI, Russian Federation; JINR; MESTD, Serbia; MSSR, Slovakia; ARRS and MIZ\v{S}, Slovenia; DST/NRF, South Africa; MINECO, Spain; SRC and Wallenberg Foundation, Sweden; SERI, SNSF and Cantons of Bern and Geneva, Switzerland; MOST, Taiwan; TAEK, Turkey; STFC, United Kingdom; DOE and NSF, United States of America. In addition, individual groups and members have received support from BCKDF, the Canada Council, CANARIE, CRC, Compute Canada, FQRNT, and the Ontario Innovation Trust, Canada; EPLANET, ERC, FP7, Horizon 2020 and Marie Skłodowska-Curie Actions, European Union; Investissements d'Avenir Labex and Idex, ANR, Region Auvergne and Fondation Partager le Savoir, France; DFG and AvH Foundation, Germany; Herakleitos, Thales and Aristeia programmes co-financed by EU-ESF and the Greek NSRF; BSF, GIF and Minerva, Israel; BRF, Norway; the Royal Society and Leverhulme Trust, United Kingdom.

The crucial computing support from all WLCG partners is acknowledged
gratefully, in particular from CERN and the ATLAS Tier-1 facilities at
TRIUMF (Canada), NDGF (Denmark, Norway, Sweden), CC-IN2P3 (France),
KIT/GridKA (Germany), INFN-CNAF (Italy), NL-T1 (Netherlands), PIC (Spain),
ASGC (Taiwan), RAL (UK) and BNL (USA) and in the Tier-2 facilities
worldwide.

\clearpage
\appendix
\part*{Appendix}
\addcontentsline{toc}{part}{Appendix}

\section{Parameter dependence of the folding and background models}
\label{app:param_dependence}

The derivation of the efficiency, resolution, and background models, described in \Equations{eq:efficiency_model}{eq:measurement_model}, is based on the form of 
these distributions in MC simulation. For the background model, constructed from the sum of all predicted backgrounds with an appreciable effect on the 
distribution, this includes events containing top quarks, primarily from \ttbar\ production, the distribution of which is affected by changing the values of the 
couplings \vl\ and \gr. The efficiency and resolution models are averages over all unmeasured distributions in the signal. Variations in the values of \vl\ and \gr\ 
alter those unmeasured distributions, which could lead to a dependence on these couplings for the efficiency and resolution models. For instance, \tch\ single 
top-quark production depends on anomalous coupling in both the top-quark production and decay vertices, so varying the couplings alters production-side 
distributions, such as the \pt\ and $\eta$ distributions of the top or spectator quark.

\begin{figure}[!htb]
  \centering
  \centerline{\makebox{
      \subfloat[]{\includegraphics[width=3.4in]{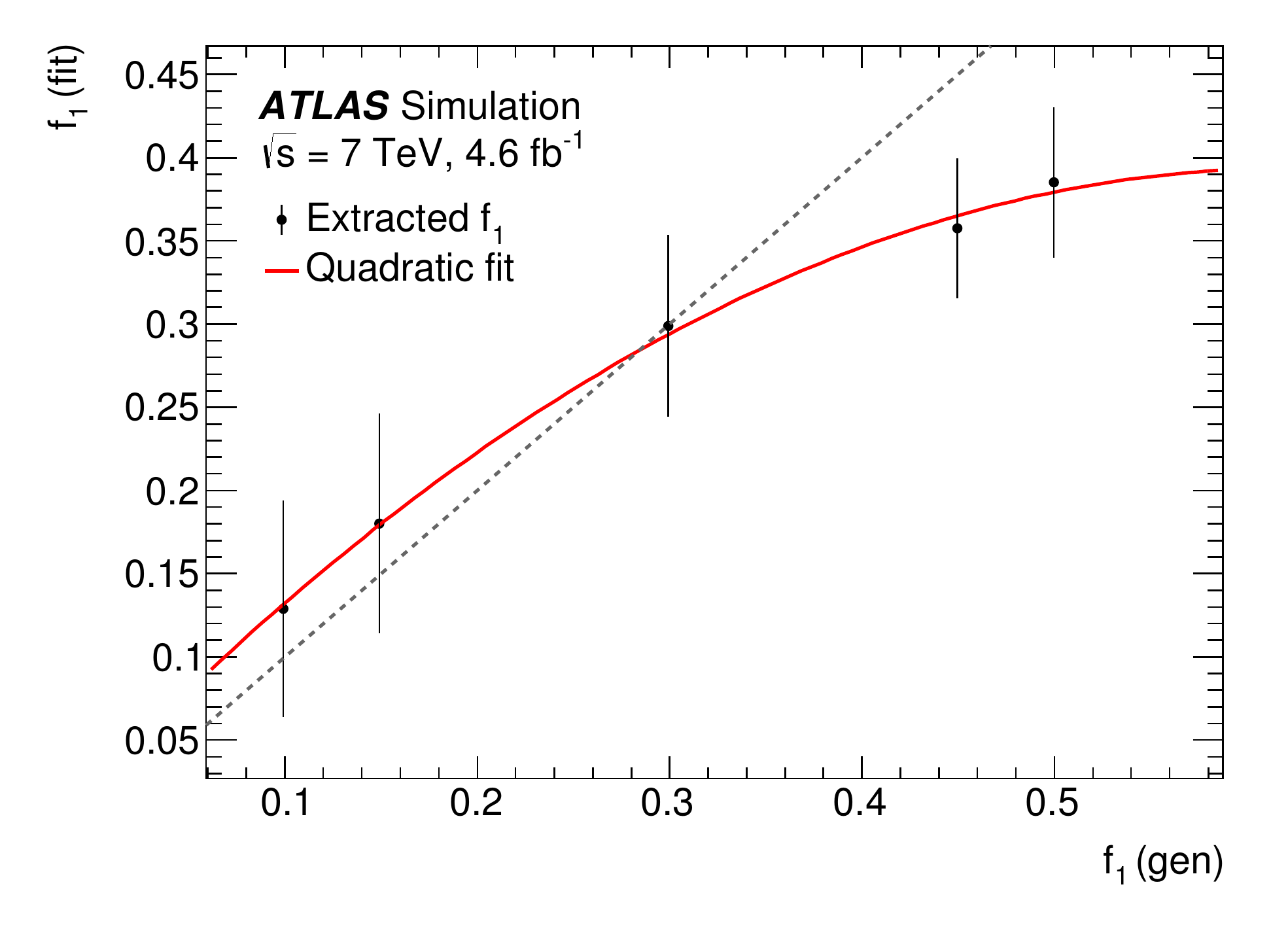}}
      \subfloat[]{\includegraphics[width=3.4in]{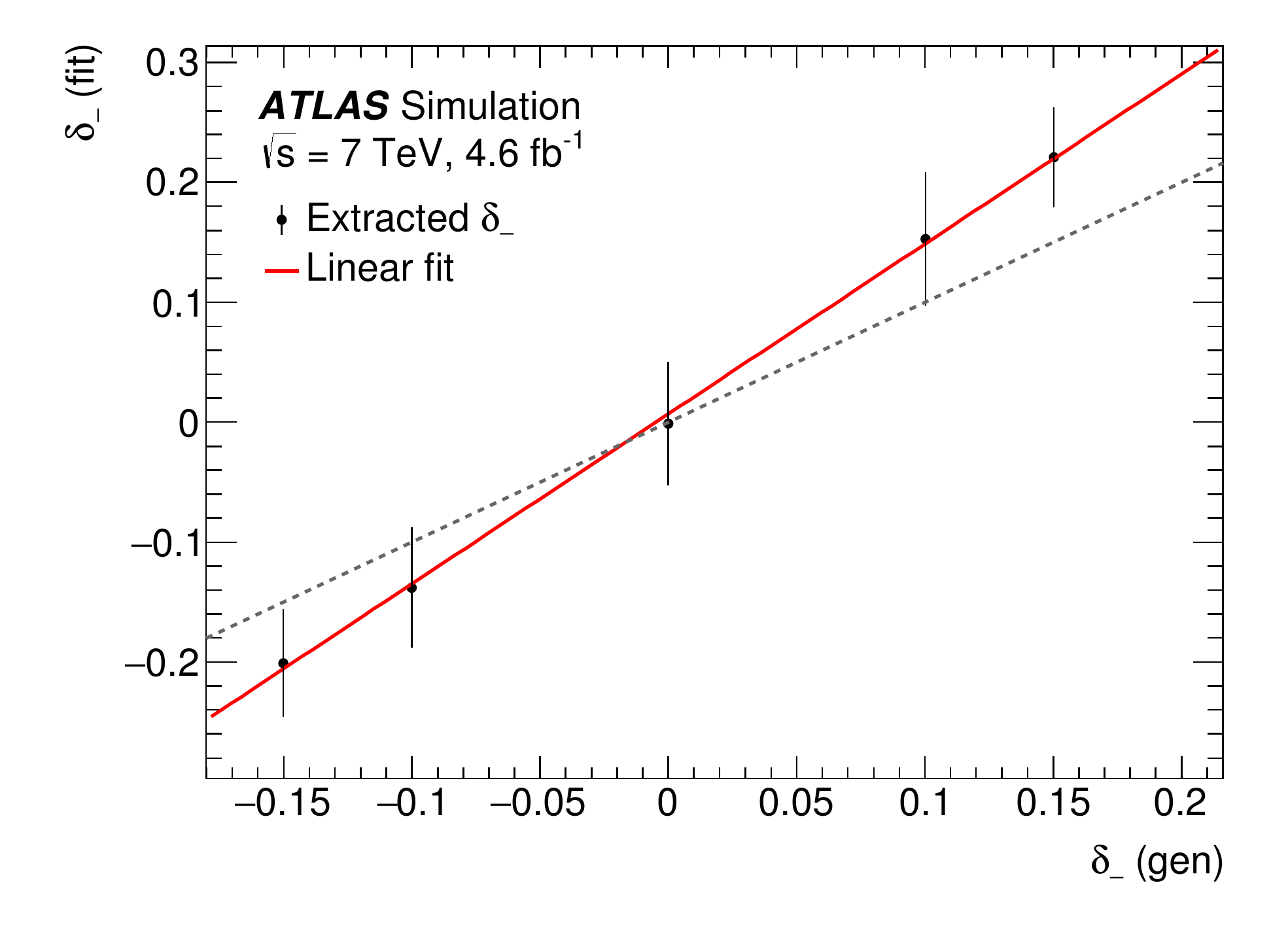}}
  }}
  \caption{ Biases in the estimation of the values of the parameters (a) $\fu$ and (b) $\delm$ which arise when the efficiency, resolution, and background models
            are determined from SM simulated events, where $\fu=0.3$ and $\delm=0$. The points represent the central values obtained from the likelihood fit, with 
            the expected statistical uncertainty in \lumi\ of data at \cmenergy\ shown by the error bars. The dashed line with a slope of one and intercept of zero, 
            representing no bias in the measurement, is added to guide the eye. A non-negligible bias is evident, and is represented at the lowest reasonable order 
            with a polynomial fit. For $\delm$ a linear fit is sufficient, while for $\fu$ a quadratic fit gives a better description. }
\label{fig:biasSMParams}
\end{figure}

To test how these potential sources of dependence affect the models, closure tests are performed with input MC events generated at different values of the 
parameters $\fu$ and $\delm$ using efficiency, resolution, and background models derived from MC events generated with SM parameter values. The \tch\ signal and 
\ttbar\ background events are generated with \protos\ for both the input MC events and the events used to derive the models. The results of these closure tests can 
then be compared to the values with which the input events were generated; any trend connecting them other than a line with a slope of one and an intercept of zero 
indicates a dependence which could bias the measurement. The results of these tests, shown in \Figure{fig:biasSMParams}, indicate biases in the extracted parameters.

\begin{figure}[!htb]
  \centering
  \centerline{\makebox{
      \subfloat[]{\includegraphics[width=3.4in]{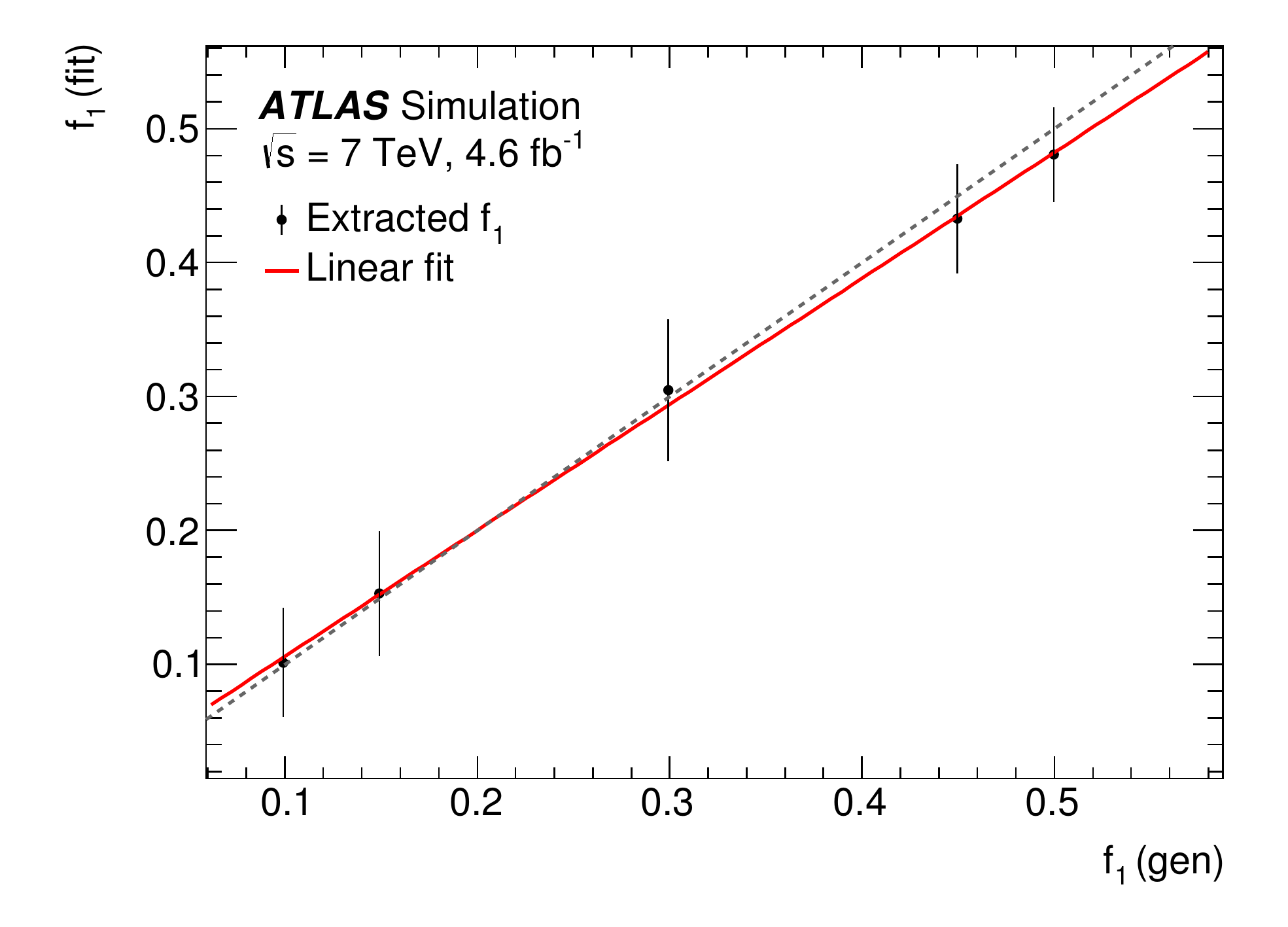}}
      \subfloat[]{\includegraphics[width=3.4in]{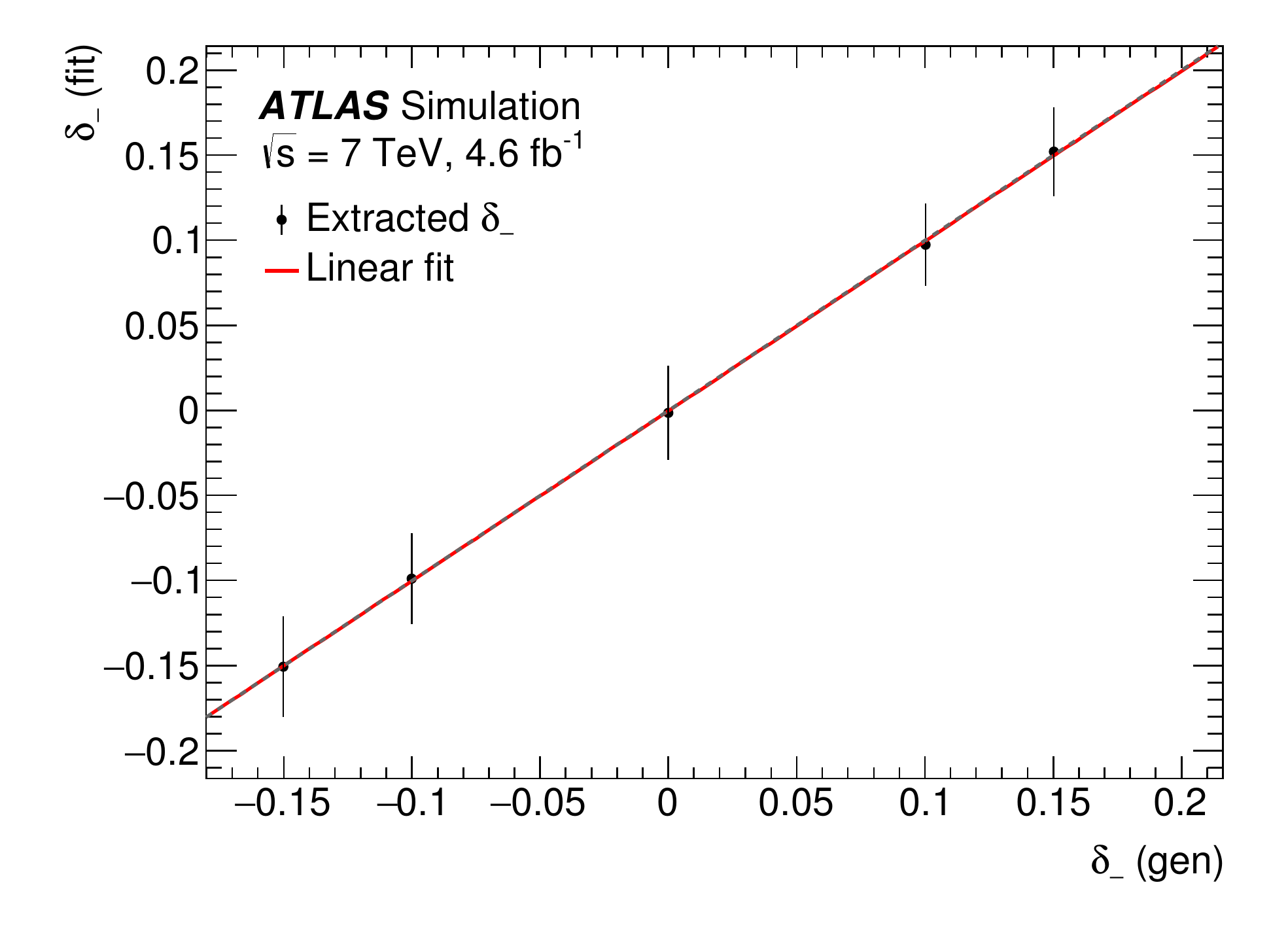}}
  }}
  \caption{ Comparison of the estimation of the values of the parameters (a) $\fu$ and (b) $\delm$ derived by allowing their values in the efficiency, resolution, 
            and background models to vary together with the values in the signal model. The points represent the central values obtained from the likelihood fit, 
            with the expected statistical uncertainty in \lumi\ of data at \cmenergy\ shown by the error bars. The dashed line with a slope of one and intercept of 
            zero, representing no bias in the measurement, is added to guide the eye. All estimates are consistent with this line; linear fits produce a slope 
            consistent with one and an intercept consistent with zero. }
\label{fig:biasRWTParams}
\end{figure}

If the biases shown are due to the mismatch between the values of $\fu$ and $\delm$ in the efficiency, resolution, and background models, they can be removed by 
varying these parameters in the functions describing these models to match their values in the signal model. This is achieved by deriving these models from \tch\ 
signal MC events representing each point in $(\fu,\delm)$ independently, and calculating the likelihood using the same $\fu$ and $\delm$ in the signal model and in 
the \tch\ signal MC events. The results of these tests, shown in \Figure{fig:biasRWTParams}, indicate that this procedure eliminates the biases shown in 
\Figure{fig:biasSMParams}. This procedure for deriving the efficiency, resolution, and background models is used in the measurement. 

\printbibliography

\newpage
\begin{flushleft}
{\Large The ATLAS Collaboration}

\bigskip

G.~Aad$^{\rm 85}$,
B.~Abbott$^{\rm 113}$,
J.~Abdallah$^{\rm 151}$,
O.~Abdinov$^{\rm 11}$,
R.~Aben$^{\rm 107}$,
M.~Abolins$^{\rm 90}$,
O.S.~AbouZeid$^{\rm 158}$,
H.~Abramowicz$^{\rm 153}$,
H.~Abreu$^{\rm 152}$,
R.~Abreu$^{\rm 116}$
% ...
\bigskip
\\
% ...
$^{11}$ Institute of Physics, Azerbaijan Academy of Sciences, Baku, Azerbaijan\\
% ...
$^{85}$ CPPM, Aix-Marseille Universit{\'e} and CNRS/IN2P3, Marseille, France\\
% ...
$^{90}$ Department of Physics and Astronomy, Michigan State University, East Lansing MI, United States of America\\
% ...
$^{107}$ Nikhef National Institute for Subatomic Physics and University of Amsterdam, Amsterdam, Netherlands\\
% ...
$^{113}$ Homer L. Dodge Department of Physics and Astronomy, University of Oklahoma, Norman OK, United States of America\\
% ...
$^{116}$ Center for High Energy Physics, University of Oregon, Eugene OR, United States of America\\
% ...
$^{151}$ Institute of Physics, Academia Sinica, Taipei, Taiwan\\
$^{152}$ Department of Physics, Technion: Israel Institute of Technology, Haifa, Israel\\
$^{153}$ Raymond and Beverly Sackler School of Physics and Astronomy, Tel Aviv University, Tel Aviv, Israel\\
% ...
$^{158}$ Department of Physics, University of Toronto, Toronto ON, Canada\\
% ...
\end{flushleft}



\end{document}